\documentclass{mn2e}
\usepackage{graphicx,epsfig}
\def\EE#1{\times 10^{#1}}

\def\cm3{\rm ~cm^{-3}}
\def\cm2{\rm ~cm^{-2}}
\def\kms{\rm ~km~s^{-1}}

\def\wll{~\lambda\lambda}

\def\Ha{{\rm H}\alpha}

\def\Msun{~{\rm M}_\odot}

\def\Ti44{M(^{44}{\rm Ti})}

\def\Msunyr{~{\rm M}_\odot~{\rm yr}^{-1}}
\def\Mdot{\dot M}

\def\lsim{\!\!\!\phantom{\le}\smash{\buildrel{}\over
  {\lower2.5dd\hbox{$\buildrel{\lower2dd\hbox{$\displaystyle<$}}\over
                               \sim$}}}\,\,}

\def\gsim{\!\!\!\phantom{\ge}\smash{\buildrel{}\over
  {\lower2.5dd\hbox{$\buildrel{\lower2dd\hbox{$\displaystyle>$}}\over
                               \sim$}}}\,\,}

\title[Hydrogen and helium in the spectra of Type Ia supernovae]{Hydrogen and helium in the spectra of Type Ia supernovae}
\author[Lundqvist et al.]{Peter Lundqvist$^{1,2}$\thanks{E-mail:
peter@astro.su.se}, Seppo Mattila$^{3}$, Jesper Sollerman$^{1,2}$, Cecilia Kozma$^{1,2}$,
E. Baron$^{4}$,
\newauthor Nick L.~J. Cox$^{5}$, Claes Fransson$^{1,2}$, Bruno Leibundgut$^{6}$,
Jason Spyromilio$^{6}$\\
$^{1}$Department of Astronomy, AlbaNova Science Center, Stockholm University,
SE-106 91 Stockholm, Sweden.\\
$^{2}$The Oskar Klein Centre, AlbaNova, SE-106 91 Stockholm, Sweden.\\
$^{3}$Finnish Centre for Astronomy with ESO (FINCA), University of Turku,
V\"ais\"al\"antie 20, FI-21500 Piikki\"o, Finland.\\
$^{4}$Department of Physics and Astronomy, University of Oklahoma,
Norman OK 73019-0225, U.S.A.\\
$^{5}$Instituut voor Sterrenkunde, KU Leuven, Celestijnenlaan 200D, bus
2401, B-3000, Leuven, Belgium \\
$^{6}$European Southern Observatory, Karl-Schwarzschild-Strasse 2,
D-85748 Garching bei M\"unchen, Germany.\\
}
\begin{document}

\date{Accepted 2013 July 14 2013. Received 2013 June 30;  in original form 2013 May 10}

\pagerange{\pageref{firstpage}--\pageref{lastpage}} \pubyear{2013}

\maketitle

\label{firstpage}

\begin{abstract}
We present predictions for hydrogen and helium emission line luminosities
from circumstellar matter around Type Ia supernovae (SNe~Ia) using time dependent
photoionization modeling. Early high-resolution ESO/VLT optical echelle
spectra of the SN~Ia 2000cx were taken before and up 
to $\sim 70$ days after maximum to probe the existence of such narrow 
emission lines from the supernova. We detect no such lines, 
and from our modeling place an upper limit on the mass loss 
rate for the putative wind from the progenitor system, 
$\Mdot \lsim 1.3\EE{-5} \Msunyr$, assuming a speed of $10 \kms$ and solar 
abundances for the wind. If the wind would be helium-enriched and/or faster, 
the upper limit on $\Mdot$ could be significantly higher. In the 
helium-enriched case, we show that the best 
line to constrain the mass loss would be He~I~$\lambda 10,830$.
In addition to confirming the details of interstellar Na~I and Ca~II
absorption towards SN~2000cx as discussed by Patat et al., we also find evidence 
for 6613.56~\AA\ Diffuse Interstellar Band (DIB) absorption in the Milky Way.
We also discuss measurements of the X-ray emission from the interaction 
between the supernova ejecta and the wind and we re-evaluate observations 
of SN 1992A obtained $\sim 16$ days after maximum by Schlegel \& Petre. We 
find an upper limit of $\Mdot \sim 1.3\EE{-5} \Msunyr$ which is significantly 
higher than that estimated by Schlegel \& Petre. These results, together with the previous 
observational work on the normal SNe~Ia 1994D and 2001el, disfavour a symbiotic star in the
upper mass loss rate regime (so called Mira type systems) from being the likely progenitor scenario 
for these SNe. Our model calculations are general, and can also be used for the subclass of
SNe~Ia that do show circumstellar interaction, e.g., the recent PTF 11kx.
 To constrain hydrogen in late time spectra, we present ESO/VLT 
and ESO/NTT optical and infrared observations of SNe~Ia 1998bu and 2000cx 
in the nebular phase, $251-388$ days after maximum. 
We see no signs of hydrogen line emission in SNe~1998bu and 2000cx at these epochs, 
and from the absence of H$\alpha$ with a width of the order $\sim 10^3 \kms$, we argue from
modeling that the mass of such hydrogen-rich gas must be $\lsim 0.03 \Msun$ for both supernovae. 
Comparing similar upper limits with recent models of Pan et al., it seems hydrogen-rich donors with a 
separation of  $\lsim$5 times the radius of the donor may be ruled out for the five 
SNe~Ia~1998bu, 2000cx, 2001el, 2005am and 2005cf. Larger separation,
helium-rich donors, or a double-degenerate origin for these supernovae seems more
likely. Our models have also been used to put the limit on hydrogen-rich gas in the recent  SN 2011fe, 
and for this supernova, a double-degenerate origin seems likely. 
\end{abstract}


\section{Introduction}
The origin of Type Ia supernovae (SNe~Ia) in general is still unknown. While it is widely accepted
that they result from the explosion of a white dwarf in a binary system, 
we are ignorant about the nature of the companion star. Branch et al. (1995)
list possible types of systems, and argue that the most likely system is a
C-O white dwarf (Hoyle \& Fowler 1960) which accretes matter from the
companion, either through Roche lobe overflow (Whelan \& Iben 1973), or as a
merger with another C-O white dwarf (Webbink 1984; Iben \& Tutukov 1984;
Paczy\'nski 1985). The natural total mass to excede to cause an explosion is
the Chandrasekhar mass, although sub-Chandrasekhar mass models have also been
proposed (Taam 1980). For a more comprehensive discussion on the different
scenarios from observational and modeling points of view we refer to Branch
et al. (1995), Branch (1998) and Hillebrandt \& Niemeyer (2000).

The current uncertainty of the origin of SNe~Ia is quite embarrassing
considering their important role in today's mapping of the geometry and
acceleration of the Universe (e.g., Schmidt et al. 1998; Perlmutter et al.
1999), as well as their huge impact on chemical evolution (e.g., Nomoto et al.
1999). To reveal the true nature of SNe~Ia, and make more accurate use of
them, we therefore need stringent methods to discriminate between possible
progenitor scenarios. An interesting method is to search for a surviving
companion star in a Type Ia supernova remnant. The detection of a binary
companion in the Tycho supernova remnant by Ruiz-Lapuente et al. (2004) is
still controversial. Schaefer \& Pagnotta (2012) recently reported a
non-detection of such a companion in the central region of the supernova 
remnant 0509-67.5 in the Large Magellanic Cloud, ruling out all single
degenerate models for a SN~Ia responsible for that remnant (see also Shappee et al. 2013a). 
Modeling of the supernova remnants interacting with possible circumstellar gas far 
away from the explosion site (e.g., Chiotellis et al. 2012) is mainly limited to 
a few historical SNe. For the recent SN 2011fe in the nearby 
($\sim$ 6.4 Mpc) galaxy M101, for which archival pre-explosion images 
from the Hubble Space Telesope (HST) were available, 
Li et al. (2011) placed an upper limit on the luminosity of the progenitor system. 
They found that a companion star with a mass above 3.5 M$_{\odot}$ could be 
ruled out including luminous red giants and almost all helium stars.

In all non-merging scenarios a wind from the companion star would be expected.
The density of this wind depends on the geometry of the mass transfer, as
well as of the mass loss rate of the companion and the wind speed. If the
wind is ionized and dense enough, it could reveal itself in the form of 
narrow emission lines before being overtaken by the supernova blast wave, 
just as in narrow-line core-collapse supernovae, SNe~IIn. If hydrogen dominates 
the wind, H$\alpha$ would be emitted (e.g., Cumming et al. 1996, henceforth
C96), and if  helium dominates He~I~$\lambda\lambda$5876, 10,830 and 
He~II~$\lambda$4686 may be prominent. The helium-dominated case is 
particularly interesting for the evolutionary path leading to a SN~Ia as
worked out by Hachisu et  al. (1999a, 2008). In that scenario the progenitor
system is a helium-rich super-soft X-ray source. Left-over material from a
merger of two white dwarfs may also create a wind-like surrounding (e.g.,
Benz et al. 1990), but is likely to be less extended, and will not contain
hydrogen and helium. 

One could also expect to see absorption lines from a circumstellar medium (CSM). 
The strength of these, however, depend strongly on the ionizing radiation from
the supernova. For a situation similar to core-collapse SNe one could expect 
high-ionization lines like C~IV~$\lambda1550$, N~V~$\lambda1240$ or 
O~VI~$\lambda1034$ (e.g., Lundqvist \& Fransson 1988), but if the mass loss 
from the companion occurs more in forms of episodic mass ejections, there 
might be very little circumstellar gas in the vicinity of the explosion, 
and there would be no circumstellar interaction boosting the ionizing 
radiation from the supernova. The mass lost from the companion could in that
case be close to neutral. Patat et al. (2007b) report time-varying Na~I~D 
absorption in spectra of SN~2006X and interpret this as evidence of 
circumstellar gas around the supernova. No similar change in absorption was seen in 
Ca~II~H\&K which puts important limits on ionizing radiation. According to 
their analysis the circumstellar shell around SN~2006X would be too faint in, 
e.g., H$\alpha$ to be detected. Chugai (2008) has reanalyzed the 
data for SN 2006X and concludes that the spectral features Patat et al. see 
may not be circumstellar but rather a geometrical effect. The latter could be 
supported by the likely absence (Crotts \& Yourdon 2008) rather 
than presence (Wang et al. 2008) of circumstellar dust around SN 2006X,
as well as absence of radio emission two years after the explosion (Chandra
et al. 2008). SN~2006X is, however, not alone to show time-varying line
absorption. SN~2007le shows similar variations, and in a statistical study
of absorbing material towards 35 Type Ia SNe, Sternberg et al. (2011)
interpreted their findings as evidence for circumstellar gas that was ejected
by the progenitor system prior to the explosion. More recently, Patat et al.
(2011) studied the interstellar absorption lines towards the recurrent nova
system RS Oph and found strong similarities with the absorption line components
detected towards SN~2006X suggesting a link between RS Oph and the progenitor
system of SN~2006X.

So, while evidence for intrinsic line absorption in SNe~Ia may be piling
up, no clear evidence of narrow emission lines have yet been seen in spectra of 
normal SNe~Ia. SN~2002ic, which could have been a SN~Ia disguised as a SN~IIn, 
was certainly a strong narrow-line emitter of mainly Balmer lines (Hamuy et
al. 2003; Kotak et al. 2004), but it is not yet clear which is the origin of
this supernova. SN~2005gj is similar to SN~2002ic (Prieto et al. 2005;
Aldering et al. 2006), and it has also been argued that both SN~1997cy and
SN~1999E could be members of the same class of objects (e.g., Hamuy et al.
2003; Deng et al. 2004; Chugai \& Yungelsson 2004; Chugai et al. 2004). Chugai
et al. (2004; see also Chugai \& Yungelson 2004) argue that the most likely
origin of SN~2002ic is a so-called Type 1.5 supernova which does not need mass
transfer from a companion (see Iben \& Renzini 1983), and that this class of
objects only makes up of order $\sim 1$\% of all SNe~Ia, and therefore does
not reveal the origin of the vast majority of SNe~Ia used for, e.g.,
cosmological studies. There are, however, other interpretations of these
objects such as a double-degenerate system (Livio \& Riess 2003), a
single-degenerate star with a massive donor (Han \& Podsiadlowski 2006) or
perhaps that these SNe are not SNe~Ia at all but instead SNe~Ic (Benetti et
al. 2006). Furthermore, Trundle et al. (2008) presented an alternative
scenario of SN 2005gj being linked to a luminous blue variable progenitor.
More recently, Dilday et al. (2012) presented a spectroscopic sequence of
PTF 11kx evolving from a SN~1999aa like SN~Ia to a spectrum dominated by
strong H${\alpha}$ emission and resembling SN 2002ic. Even more recently,
Silverman et al. (2013) discussed the nature of a total of 16 events showing an
underlying spectrum resembling a SN~Ia and being dominated by relative narrow
($\sim$2000 km s$^{-1}$) H${\alpha}$ emission line from CSM interaction.
Although, the fraction of SNe~Ia showing a prominent CSM interaction is
pretty low, $\sim$0.1-1\% (Dilday et al. 2012), these studies demonstrate that
at least some SNe~Ia could originate from a single degenerate progenitor system
resulting in a hydrogen-rich CSM within which the SN explodes. An alternative
to nova-like shells (Dilday et al. 2012) is that the CSM structure is formed by a 
merger of a white dwarf with the core of a massive asymptotic giant branch companion, 
ending their common envelope phase and ejecting multiple shells (Soker et al. 2013).


In this paper we concentrate on constraining hydrogen and helium in more
normal SNe~Ia. Our approach is to study what emission would occur when the
supernova ejecta interact with circumstellar gas already shortly after the
explosion. A first attempt to observe and model this situation was done by
C96 for SN 1994D. The spectrum was obtained $\sim 10$ days before maximum and 
the analysis involved full time-dependent photoionization calculations to estimate the
narrow H$\alpha$ emission from the tentative wind. The somewhat refined
analysis in Lundqvist \& Cumming (1997; henceforth LC97) gives an upper limit
on the mass loss from the progenitor system of $\Mdot \lsim 1.5\EE{-5}
\Msunyr$. This limit assumed solar abundances for the wind and a wind speed
of $v_{w} = 10 \kms$. 

C96 argued that observations as early as possible are needed to detect circumstellar
gas in emission because the densest part of the wind may quickly be overtaken by 
the expanding supernova ejecta. In Mattila et al. (2005; henceforth M05) we
presented the results of our VLT/UVES observations of SN~2001el at 9 and 2
days before maximum. These yielded a mass loss rate upper limit of
$\Mdot \lsim 9\EE{-6} \Msunyr$ assuming a wind velocity of 10 $\kms$.
Here we report on another supernova observed within this program, SN 2000cx
(Sect. 3.1). It was suitably located in the outskirts, 23\farcs0 west and
109\farcs3 south of the S0 galaxy NGC 524. The host galaxy has a recession
velocity of $2353-2379 \kms$ (Simien \& Prugniel 2000, Emsellem et al. 2004), 
and the galactic reddening in the direction of NGC 524 is modest, $E(B-V)=0.083$ 
(Schlegel et al. 1998). A low dispersion spectrum obtained on July 23 revealed this
supernova to be of Type Ia (Chornock et al. 2000), resembling the
overluminous SN 1991T a few days before maximum (see also Li et al. 2001).
SN 2000cx continued to rise, in a similar fashion to SN 1994D (Li et al. 2001), 
and reached maximum $B$ brightness ($B \approx 13.4$) on July 26.7 (Li et al.
2001). However, the postmaximum decline in $B$-band was slower than for
SN 1994D, whereas the $V$, $R$ and $I$ light curves declined faster than for
most other SNe~Ia (Li et al. 2001). 

In our analysis of SN 2000cx we discuss both interstellar lines and limits on 
circumstellar emission lines. Preliminary results for SN 2000cx were already presented
in Lundqvist et al. (2003; 2005). Some analysis of the data is also presented by
Patat et al. (2007a) to check whether the supernova displayed time-varying line
absorption similar to that of SN~2006X. No such variation was found by these
authors. In our analysis, we have also calculated refined photoionization
models (Sect. 2) compared to the earlier models in C96 and LC97. In Sect. 3 and 4
we discuss the results and compare them with constraints on circumstellar interaction in
SNe~Ia from observations at other wavelengths. 
  
We also present ESO-VLT and ESO New Technology Telescope (NTT) observations
of SN~2000cx (Sect. 3.1.3) and SN~1998bu (Sect. 3.2) taken 363 and $251-388$
days after $B$-band maximum, respectively, i.e., in the nebular phase.
According to Marietta et al. (2000) this is the epoch when a companion star
in the single-degenerate scenario could reveal its hydrogen-rich envelope.
Kasen (2010) suggest that early time photometry could be
used to detect signatures of the collision of the SN ejecta with its companion
star. The impact of SN~Ia ejecta on a non-degenerate binary companion has
been recently invistigated also by Pan et al. (2012) and Liu et al. (2012).
In the models of Marietta et al. the atmosphere of a close companion 
would be lost as a result of the supernova explosion. Apart from some mass
being stripped away at the impact, the largest fraction of the gas lost will 
stream away at a velocity of $\lsim 10^3 \kms$. Heated by radioactive decays 
it could emit lines of hydrogen, and possibly helium, characterized by this 
velocity. Marietta et al. highlight the possibility to detect Pa$\beta$ as it
sits in a spectral region which is rather unblended by other lines. Pa$\beta$ 
is also more accessible from the ground than, e.g., Pa$\alpha$. However, as 
we will show from our modeling in Sect. 2, H$\alpha$ is actually a more 
promising probe than any of the IR lines. In any case, detecting such lines 
gives a possibility to test the non-merging scenario even in the case a 
progenitor wind would be too dilute to be detected. Similar studies based on
late time H$\alpha$ emission were presented for SN 2001el in M05 for
SNe 2005am and 2005cf in and Leonard (2007). More recently, Shappee et al.
(2013b) presented late time spectroscopy of SN~2011fe and obtained the lowest
yet upper limit for the amount of stripped material at SN~Ia explosion
site. All these studies rest on the theoretical framework presented here.
Furthermore, the lack of radio emission (Chomiuk et al. 2012) implies an extremely
low mass loss rate from the progenitor system of SN~2011fe. While there
are indeed single degenerate scenarios where very low CSM densities are expected (e.g., 
Di Stefano et al. 2011, Justham 2011), a possible problem for the single degenerate
explanation for SN~2011fe is the lack of signatures from a companion (Shappee et al. 2013a).

Some recent SNe~Ia have shown that their broad photospheric Ca~II supernova 
line profiles have high-velocity extensions that look like the signature
of a high-velocity shell-like structure. The high-velocity stucture have
been discussed for SN~2000cx itself by Thomas et al. (2004) and Branch et 
al. (2004). They have also been observed to be present in a number of other
SNe Ia (see Mazzali et al. 2005). Although the reason for the high-velocity 
shell-like structure is still unknown, Gerardy et al. (2003) argue that the 
evolution of the profile of the Ca II IR triplet in SN~2003du indicates early 
circumstellar interaction. In Sect. 3.3 we briefly discuss the photospheric broad Ca~II 
supernova line profiles in our early data of SN~2000cx. (A more extended 
discussion on early high-velocity features of SNe~Ia was provided in our 
paper on SN~2001el by M05. See also the comprehensive study by Tanaka et al.
2008). We summarize our results and discussions for both early and late
phases in Sect. 4.

\section{Model calculations and Results}
\subsection{Circumstellar interaction}
To obtain an upper limit on the density of the putative circumstellar gas around 
SN 2000cx and other SNe~Ia, we have modeled the line emission in a way similar to 
C96 and LC97, although with some important refinements.
We assume that the supernova ejecta have a density profile 
of $\rho_{ej} \propto r^{-7}$ (like in the W7 model of Nomoto et 
al. 1984), and that the ejecta interact with the wind of a binary companion 
which has a density profile of $\rho_{w} \propto r^{-2}$. 
Very early, i.e., during the first few hours after explosion, the radius of 
the outermost supernova ejecta is comparable to the presumed binary 
separation, and the ejecta expand into an asymmetric 
environment. But already after $\sim 1$ day the interaction between the
ejecta and the wind occurs sufficiently far away from the center of 
explosion and the center of wind emission to justify that the interaction
can be treated as point-symmetric. For simplicity, we will assume
a spherically symmetric situation. 

The assumption of power-law density distributions for the ejecta and the wind
makes it possible to use similarity solutions for the expansion and structure
of the interaction region (Chevalier 1982a). This is obviously a simplification 
since the wind structure close to the white dwarf is uncertain, and the 
density profile of the outer ejecta may be quite different
from $\rho_{ej} \propto r^{-7}$ early on (e.g., Branch et al. 1985), 
possibly closer to an exponential fall-off (Dwarkadas \& Chevalier 1998). 
Ideally, one should therefore model the interaction of 
the ejecta with the circumstellar gas numerically  rather than using similarity 
solutions. However, considering that we are only interested in the evolution 
after $\sim 1$ day, when the inner region of the wind has been overtaken,
and part of the outer ejecta has been shocked, we have followed the method 
of C96 to use similarity solutions to describe the expansion and structure 
of the interaction region. 

We start our calculations at $t_0=1.0$~day after explosion, and at this epoch 
we assume that the maximum velocity of the ejecta is $V_{ej} = 4.5\EE4 \kms$.
This is higher than in C96, but corresponds better to the fastest ejecta in 
explosion models (see, e.g., Branch et al. 1985), as well as the highest 
velocity seen in Ca~II for SNe~1984A, 1990N, 2001el (M05) 
and SN~2003du (Stanishev et al., 2007) about 10 days before 
maximum, $V_{ej} > 3\EE4 \kms$ (see also Sect. 3.3 and the compilation
of high-velocity features in SN~Ia by Mazzali et al. 2005). Note the 
difference in time zero in our models, and in general for supernova light 
curves which normally refers to $B$-band maximum which occurs $15-20$ days 
after the explosion. With the adopted slopes 
of the ejecta and the circumstellar gas, the fall-off of the maximum ejecta velocity 
conforms to $V_{ej} \propto t^{-0.2}$, according to the similarity solutions 
of Chevalier (1982a). At 1 day, the velocities of the circumstellar shock and the reverse 
shock going into the ejecta 
are $V_{s} \sim 4.5\EE4 \kms$ and $V_{rev} \sim 1.1\EE4 \kms$, respectively.   

\subsubsection{Ionizing radiation}
The ionizing radiation from the interaction region is composed of 
free-free emission from the shocked ejecta and circumstellar gas, and 
Comptonized photospheric radiation in the shocked gas. 
The situation is thus similar to that discussed by Fransson (1984) and 
Lundqvist \& Fransson (1988) for SNe~1979C and 1980K, and by 
Fransson et al. (1996) for SN~1993J. The main difference 
is that the expected wind density for SN~Ia progenitors is lower than 
for these core-collapse SNe, which all had wind densities in excess 
of $\Mdot \simeq 2\EE{-5} \Msunyr v_{w10}^{-1}$ (Lundqvist \& 
Fransson 1988; Fransson \& Bj\"ornsson 1998). Here, $v_{w10}$ is the 
wind velocity in units of $10~\kms$. It is also likely that the 
structure of the outer ejecta in core-collapse SNe have steeper density 
profiles than SNe~Ia, and therefore a slower deceleration of the 
shocks and a different structure of the shocked gas (Chevalier 1982a). 
The result is a higher expected temperature of the reverse shock in SN~Ia, 
which in combination with the lower wind density means that the reverse 
shock will not be radiative for SNe~Ia in our scenario even for the highest
mass loss rates in our grid ($\Mdot \simeq 2\EE{-5} \Msunyr v_{w10}^{-1}$) 
already at $\sim 1$ day; the closest we get is $t_{\rm cool} / t \sim 2$.
Because $t_{\rm cool} / t \propto t^{0.4}$ in our model (cf. Fransson et 
al. 1996) no cool shell of shocked ejecta will form.
If inverse Compton scattering in the shocked
ejecta is important for the highest wind densities (see below), a cool shell 
could form early on.

We note that Gerardy et al. (2003) in their models for SN~2003du assumed 
that the supernova may encounter very dense circumstellar material 
immediately after the explosion, and that the density of the circumstellar gas could 
be high enough for the forward and reverse shocks to become radiative.
However, as they do not quantify possible emission of ionizing radiation and
how it could affect unshocked circumstellar gas further away from the supernova, we
cannot compare the results. In our scenario, the radiation from 
the reverse shock will reach the circumstellar gas unattenuated, making the reverse
shock emission more important for the ionization of the wind at energies 
below $\sim 10$~keV than the free-free radiation from the circumstellar shock (see 
below), and we have omitted the latter in our models.

Cumming et al. (1996) and LC97 assumed a constant density
and temperature of the shocked gas. We instead take the density and 
temperature structures of the reverse shock from similarity solutions 
(Chevalier 1982a), and use these to calculate the free-free emission 
from the shocked gas. As in C96 we note that the shocked ejecta are likely 
to be composed of C and O rather than H and He, which gives a higher 
temperature. For equal number densities of C and O,
$X({\rm C}) = X({\rm O}) = 0.5$, the temperature increase is a 
factor $\sim 1.6$ compared to solar abundances. With the parameters 
chosen in our models, the temperature of the reverse shock evolves
as $T_{rev}~\sim~2.4\EE9~t_{\rm day}^{-0.4}$~K, 
where $t_{\rm day}$ is the time in days since the explosion. 

The heavier elements also boost the free-free 
emissivity, $j_{\rm ff}(\epsilon)$, for a given temperature. The combined 
effect of temperature increase and boost in $j_{\rm ff}(\epsilon)$ compared 
to a wind of solar abundance can be exemplified by a case 
with $X({\rm C}) = X({\rm O}) = 0.5$ and $T = 4.8\EE8$~K (valid just behind
the shock front at $\sim 55$~days, i.e., somewhat later than the observed
epoch of SN 1992A, see below). The corresponding temperature for the solar
abundance composition would be $3\EE8$~K. Taking the temperature difference
into account, $j_{\rm ff}({\rm 1~keV})$ is a factor
of $\sim 1.8$ higher for the metal-rich ejecta. This obviously affects the 
derived value for the mass loss rate from X-ray observations (see below).

As in C96 and LC97 we take the finite time scale of 
electron-ion equilibration in the shocked gas  into account. We have simulated this by 
calculating models where the electron 
temperature, $T_{e}$, is a factor of 2 lower than $T_{rev}$.
Unequal electron and ion temperatures are more likely in our current models
than in the models in C96 since we now use a higher, and more realistic, 
shock velocity. Assuming pure Coulomb collisions 
and $V_{rev} = 1.1\EE4~\kms$ (cf. above), electron-ion 
equilibration in the shocked ejecta occurs for a wind described by a mass 
loss rate of $\Mdot \gsim 5\EE{-6} v_{w10}^{-1} \Msunyr$. Poorly known plasma 
instabilities may give electron-ion equilibration also for lower mass 
loss rates, and we have therefore calculated models with $T_{e} = T_{i}$ 
($T_{i}$ being the ion temperature) also for low mass loss rates.

\begin{figure*}
\includegraphics[width=150mm, clip]{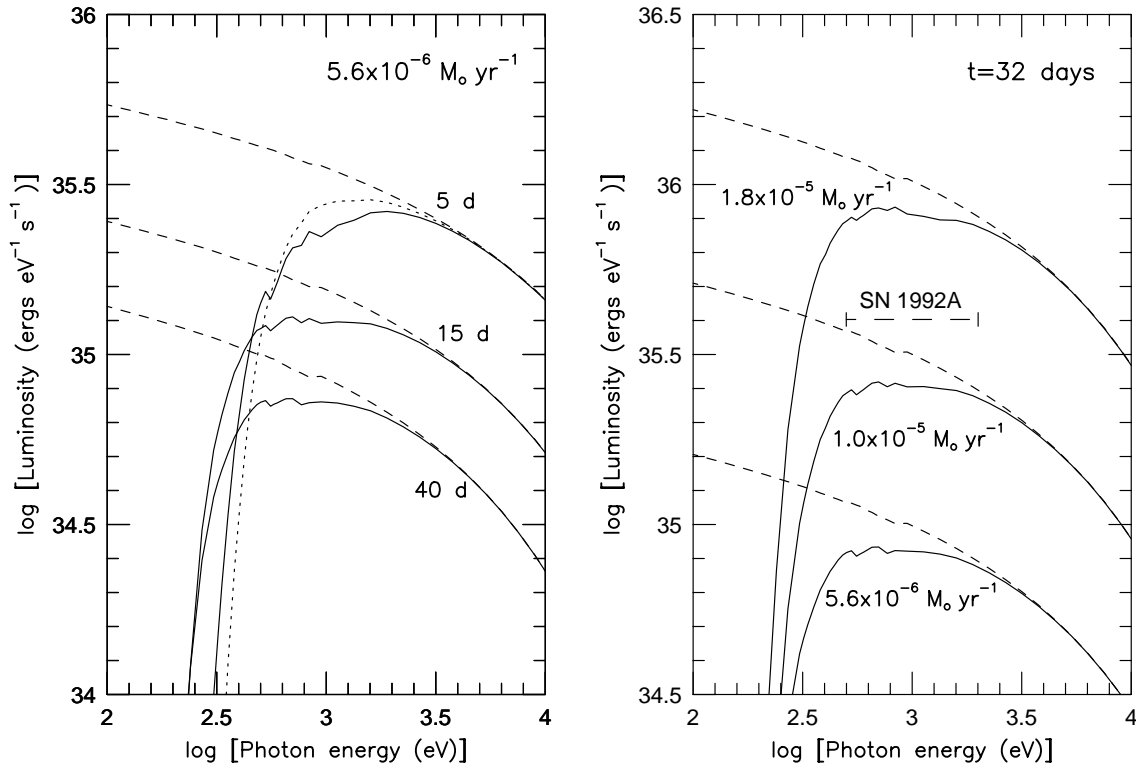}
\caption{{\bf Left:} Ionizing spectrum at 5, 15 and 40 days after the explosion
for a model with a wind density described by the mass loss 
rate $5.6\EE{-6} \Msunyr$ and wind speed $10 \kms$. Ions and electrons in the 
shocked supernova ejecta are assumed to have the same temperature.
Dashed lines show the unattenuated spectrum, while
solid lines show the spectrum after absorption in the wind. (Dotted line
shows a case for 5 days after the explosion assuming He/H = 1.0.
The other models have He/H = 0.085.) {\bf Right:} Ionizing spectrum at 
32 days after the explosion for three mass loss rates (using the same
representations as in the figure to the left). The vertical long-dashed bar
shows the upper limit on the luminosity within $0.5-2.0$ keV estimated by
Schlegel \& Petre (1993) for SN 1992A at $\sim 16$ days after maximum.
(For simplicity, we have assumed a constant flux within the bandpass used
by Schlegel \& Petre.)
}
\label{fig9_letal}
\end{figure*}

In C96 and LC97 it was assumed that the reverse shock was the sole 
source of ionizing radiation (apart from a weak preionization by the 
white dwarf). It was argued that the shock breakout and subsequent 
photospheric emission were inefficient in ionizing the wind. 
However, model calculations by Blinnikov \& Sorokina (2000) show that 
there may be a phase of ionizing radiation from the photosphere for a 
few days within the first $\sim 10$ days after the explosion. In some of
our models we have therefore included the results of Blinnikov \& Sorokina 
(2001), and in particular their w7jzl155.ph model, which is an updated version 
of the w7jz.ph model in Blinnikov \& Sorokina (2000). These models use the 
W7 model of Nomoto et al. (1984) as input, and have a peak in far-UV 
emission between days $4-6$.

For the original models (Blinnikov \& Sorokina 2000) we found that the 
photospheric emission was strong and actually dominates the ionization of 
the wind after $\sim 3$ days in our models. Including an updated UV 
line list, the models of Blinnikov \& Sorokina (2001) predict less 
far-UV emission so that it becomes less important for the ionization of 
the wind. Still, the emission from the photosphere is important for 
models with low wind densities. For example, the luminosity of ionizing 
photons emitted from the photosphere in the w7jzl155.ph model at 5~days 
is $3.0\EE{48}$~s$^{-1}$, whereas the luminosity of ionizing photons at
the same epoch from the reverse shock for a model 
with $\Mdot = 5.6\EE{-6}~v_{w10}^{-1} \Msunyr$ 
and $T_{e} = T_{i}$ is $1.6\EE{48}$~s$^{-1}$. It is obvious that 
the photospheric emission has to be modeled in even greater detail to 
investigate its far-UV emission as it could be important for the ionization 
of the surrounding medium, in particular if the circumstellar gas would
reside far out from the explosion site as in the shell model for SN~2006X
proposed by Patat et al. (2007b). In such a case the photospheric 
emission would be the sole ionizing source.

In the models of Blinnikov \& Sorokina the photospheric spectrum falls
steeply with photon energy above the ionization threshold of hydrogen, and 
is unimportant at X-ray energies. The photons emitted from the 
photosphere will, however, be boosted in energy by interaction with 
the hot electrons in the shocked ejecta and circumstellar gas. The 
shocked circumstellar gas is unlikely to have equal ion and electron temperatures, 
but it seems reasonable to assume that the electron temperature can attain a
temperature of order $10^9$ K, just as indicated for SN~1993J
(Fransson et al. 1996), especially for models in the upper range of
the values for $\Mdot/v$ we have considered. As long as this is true, 
the free-free emission from the circumstellar shock is unimportant for the ionization 
radiation compared to that from the reverse shock (because of the, in our
scenario, $\sim 50\%$ larger mass of shocked ejecta compared to shocked 
circumstellar gas as well as $\sim 6$ times higher density), but for our assumed 
ejecta velocity and density structures of the wind and ejecta, the optical 
depth to electron scattering in the shocked circumstellar 
gas is $\tau_{e} \sim 4.5\EE{-3}~\Mdot_{-6}~v_{w10}^{-1}~t_{\rm day}^{-0.8}$,
and this may create an important power-law tail to the photospheric
spectrum (cf. Fransson et al. 1996), especially at early times. 
Here $\Mdot_{-6}$ is the mass loss rate of the wind in units
of $10^{-6} \Msunyr$.  
For lower values of $\Mdot/v$ Coulomb heating may not be enough to heat
the shocked wind up to $10^9$ K which will reduce the importance of
inverse Compton scattering rapidly. The free-free emission from the shocked
circumstellar gas could in such cases instead become more important as it would 
increase in strength at energies below $\sim 1$ keV. Early on, inverse
Compton scattering of the shocked ejecta may also be important, i.e., when
the shocked ejecta are heated up to $\sim 10^9$ K. However, already at 
10 days, the shocked ejecta temperature has fallen below $10^9$ K,
and the electron temperature may be even lower (cf. above).
Considering the uncertainty of the temperature of the forward shock,
we have omitted the ionizing radiation from the circumstellar shock.  

Figure~1 (left part) shows the spectrum of 
the ionizing radiation at the epochs 5, 15 and 40 days for a wind density 
described by $5.6\EE{-6}~v_{w10}^{-1}~\Msunyr$. The spectrum of
the {\it emitted} radiation is shown by dashed lines, 
assuming $T_{e} = T_{i}$ in the shocked ejecta. We have used the plasma
code discussed in Sorokina et al. (2004) and Mattila et al. (2008) for these
calculations. The emission is totally dominated by free-free emission (i.e., there is no line 
emission). The decrease in shock temperature scales as $\propto t^{-0.4}$, and can be 
seen as a steepening of the spectrum for $\epsilon \gsim 3$~keV with time. At lower
photon energies the luminosity decreases as $L(\epsilon) \propto t^{-0.6}$.
In C96 it was assumed that $L(\epsilon) \propto {\rm exp}(-\epsilon/kT_{rev,e})$, 
where $T_{rev,e}$ is the electron temperature of the shocked ejecta
immediately behind the shock front. This produces a flat spectrum at
energies $\epsilon \ll T_{rev,e}$. As can be seen in Fig.~1, a
more realistic structure of the shocked ejecta, with the temperature
decreasing outward according to the $n=7$ and $s=2$ model of Chevalier 
(1982a), the photon energy spectrum follows $L(\epsilon) \propto \epsilon^{-0.17}$
below $\sim 1$ keV. However, the {\it observed} spectrum seen by 
an external viewer (solid lined) is different since most of the emission 
is absorbed by the circumstellar gas at these energies. 
(See Sect. 2.1.2. for more about the ionization of the circumstellar gas.) 
In addition, absorption in the interstellar gas will further decrease 
the observed flux at low photon energies. The circumstellar absorption
is negligible only above $\sim 3$~keV. The effect of the composition of 
the circumstellar material on the far-UV and X-ray absorption is minor 
as can be seen for the flux at 5 days, where the solid line in Fig.~1 
is for H/He~=~0.085 (by number), whereas the dotted line is for H/He~=~1.0.

Figure~1 (right part) shows the spectrum (emitted and observed) 
at 32 days for three wind densities. We have chosen this epoch
since it corresponds to the epoch of observation for SN~1992A 
(Schlegel \& Petre 1993). Because the spectrum is dominated by 
free-free emission, the emitted spectrum has the 
dependence $L(\epsilon) \propto (\Mdot/v_{w})^2$. At energies
below $\sim 3$~keV absorption in the circumstellar gas becomes
important, but the $(\Mdot/v_{w})^2$ dependence holds also for 
the observed flux down to roughly half a keV at this epoch. 
In Fig.~1 we have included the observed upper limit derived
for SN~1992A (Schlegel \& Petre 1993). Assuming that our model can
be applied for SN~1992A, we obtain an upper limit on the mass loss rate
which is $\sim 1.3\EE{-5} v_{w10}^{-1} \Msunyr$. This
is substantially higher than derived by Schlegel \& Petre (1993) who
obtained $\lsim 3\EE{-6} v_{w10}^{-1} \Msunyr$. 
A reason for the discrepancy between our result and that of Schlegel \& 
Petre, except for our updated estimate of the X-ray spectrum, is
that Schlegel \& Petre used the time since supernova maximum, instead
of the time since explosion in their luminosity estimate. Their
limit is therefore clearly too low if the reverse shock dominates
the ionizing radiation. A point of caution here is that inverse Compton
scattering could be important if the electrons of the shocked circumstellar gas 
are hot enough. For $\Mdot/v_{w} \sim 1.3\EE{-5} v_{w10}^{-1} \Msunyr$, the
optical depth in the shocked circumstellar gas $\tau_{e} \sim 0.002$. Depending on
the electron temperature, this may be enough to affect our estimated
upper limit of $\Mdot/v_{w}$ for SN 1992A. Including only the free-free 
emission from the reverse shock thus gives a conservative limit.

The model presented here for the ionizing radiation from the reverse
shock was also included in the analysis of the X-ray observations of 
the normal SN~Ia~2002bo by Hughes et al. (2007), along with modeling only
assuming Coulomb heating of the electrons in the shocked gas.
The observed upper limit to the X-ray emission from SN~2002bo 
was found to correspond to $\Mdot/v_{w} \sim 2\EE{-5} v_{w10}^{-1} \Msunyr$,
i.e., slightly higher than our limit for SN~1992A. 

\subsubsection{The circumstellar gas}
The ionizing radiation produced by the shocked gas ionizes and heats the 
unshocked circumstellar gas. We have used an updated version of our time dependent 
photoionization code described in Lundqvist \& Fransson (1996), to 
calculate the temperature and ionization structure of this gas. As in C96, we have 
performed calculations for wind densities 
between $10^{-7} v_{w10}^{-1}$ - $2\times 10^{-5} v_{w10}^{-1}$ $\Msunyr$.
We have considered models both with and without a contribution of ionizing 
photons from the photosphere (cf. Sect. 3.1.1.) and we have set 
the ratio $T_{e}/ T_{rev}$ of the reverse shock to either 0.5 or 1 to study
the effect of departure from $T_{e} = T_{i}$. We have also varied the 
He/H$-$ratio of the circumstellar gas to see under which conditions circumstellar helium lines may
be more prominent than lines from hydrogen.
The results from the models presented in this paper have already been used
in M05 for SN~2001el. In that paper we only considered
models with $T_{e} = T_{i}$ for the reverse shock.

Assuming that free-free emission from the reverse shock dominates
the ionization of the circumstellar gas, the ionized part of the wind will be 
confined to a volume set by the shock radius, $R_{s}$, and an outer 
``Str\"omgren" radius, $R_{HII}$, which is only a few times larger
than $R_{s}$. The situation looks like that depicted in Fig. 1 of
Lundqvist \& Fransson (1988). For example, for a wind 
with $5.6\times 10^{-6} v_{w10}^{-1} \Msunyr$, $T_{e} = T_{i}$ of the reverse 
shock and He/H = 0.1 of the circumstellar gas, $R_{HII} / R_{s} \sim 3$ around 1 day 
after explosion. Until 10 days, $R_{HII}$ has advanced a bit less rapidly 
than $R_{s}$ so that $R_{HII}/R_{s} \sim 2.5$. The ratio of radii continues 
to decrease so that it is $\sim 1.5$ by day 50. 
For a model with $T_{e} = T_{rev} /2$, $R_{HII}/R_{s}$ is $\sim 4$, $\sim 3$ and 
$\sim 2$ at 1, 10, and 50 days, respectively. The same trend can also be seen 
for other values of $\Mdot/v_{w}$. The ionized region is slightly larger for 
models with the softer spectrum in $T_{e} < T_{i}$ models compared to model 
with $T_{e} = T_{i}$. The reason is that a slightly larger number of ionizing 
photons at lower energies are produced in the $T_{e} < T_{i}$ models. These 
photons are also more easily absorbed by the circumstellar gas as X-ray cross sections 
of ions decrease with photon energy. The main difference between models with 
different values of $\Mdot/v_{w}$ is that $R_{HII} / R_{s}$ increases with 
increasing $\Mdot/v_{w}$. For example, at 10 days $R_{HII} / R_{s} \sim 6$
for $\Mdot/v_{w} =  10^{-5} v_{w10}^{-1} \Msunyr$ and $T_{e} = T_{i}$.

\begin{figure}
\begin{center}
\includegraphics[width=55mm, clip, angle=90]{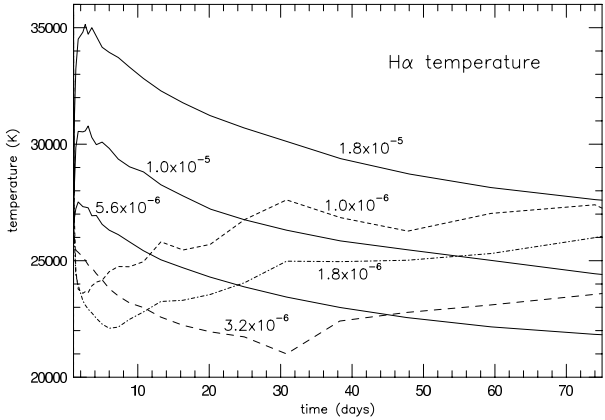}
\end{center}
\caption{
Calculated temperature of the circumstellar gas emitting H$\alpha$
as a function of time from the explosion.
(The temperature at each epoch was weighted by the luminosity of H$\alpha$.) 
We have assumed $T_{e} = T_{i}$ behind the reverse shock, no photoionizing 
radiation from the photosphere, and He/H = 0.10 for the circumstellar
gas. Models were made for different mass loss rates of the progenitor 
system, as marked in the figure in units of $\Msunyr$ for a wind speed 
of $10 \kms$. Although the temperature is in general higher in the models 
with high wind density, the H$\alpha$ temperature is rather insensitive to 
the wind density, and also remains rather constant with time. To avoid 
confusion, the models with the lowest mass loss rates have been
drawn with dashed and dash-dotted lines.}
\label{fig10_letal}
\end{figure}

Typical temperatures of the ionized (unshocked) circumstellar gas range 
between $2\times10^{4}$~K (close to $R_{HII}$) and of the order $10^{5}$~K
close to $R_{s}$, with the latter temperature decreasing with increasing
time as well as with decreasing $\Mdot/v_{w}$. Figure 2 shows the
evolution of the wind temperature, weighted with the H$\alpha$ luminosity. 
As one can see, this temperature remains within $2.2\EE4 - 3.4\EE4$ K 
for all our $T_{e} = T_{i}$ models. We use this finding below when estimating
the wind density around SN~2000cx.

Elements lighter than sulphur are almost fully ionized close to $R_{s}$, and a 
large fraction of the metals is ionized also outside $R_{HII}$ (defined as the radius 
where the relative fractions of protons and neutral hydrogen are equal). The reason
is the higher X-ray cross section of the metals compared to hydrogen.

\begin{figure}
\center{\includegraphics[width=87mm, clip,]{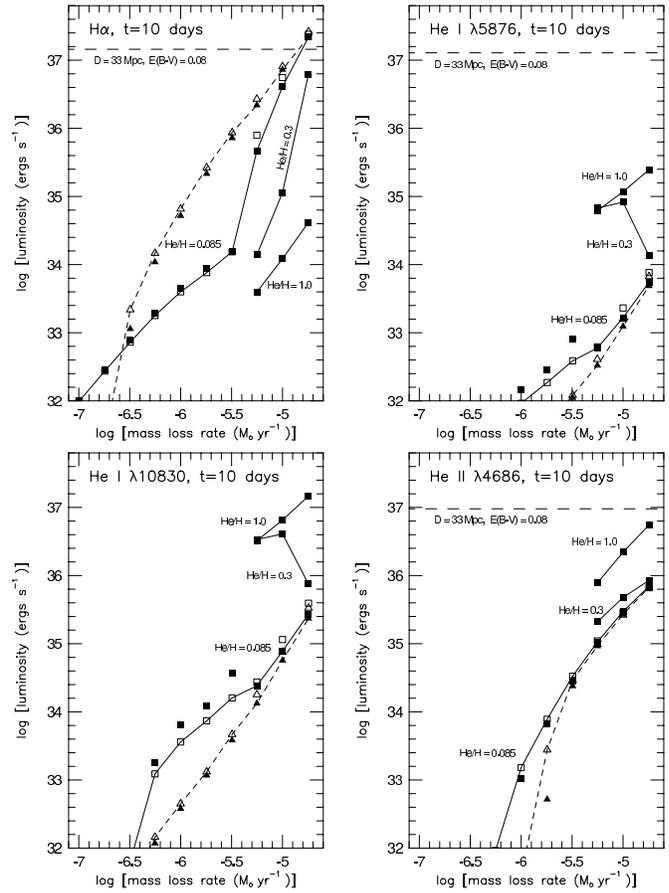}}
\caption{
Line luminosites at 10 days after the explosion as a function of mass loss rate,
assuming a wind speed of $10 \kms$. Squares show models in which ionizing
radiation is only produced by the reverse shock, while in models marked by
triangles we have also included the photospheric emission from the model
w7jzl155.ph of Blinnikov \& Sorokina (2001). Filled symbols are for temperature
equipartition between electrons and ions behind the reverse shock, whereas
for open symbols $T_{e}=T_{rev}/2$. Models are shown for three values
of the number density ratio He/H. The observed limit is for the first epoch
in Table 2 for the case of a wind temperature of $2.8\EE4$ K and a wind 
speed of $10 \kms$. For SN 2000cx we have assumed a distance of 33 Mpc and
an extinction of $E(B-V) = 0.08$. For further details, see text. 
}
\label{fig11_letal}
\end{figure}

\begin{figure}
\center{\includegraphics[width=87mm, clip,]{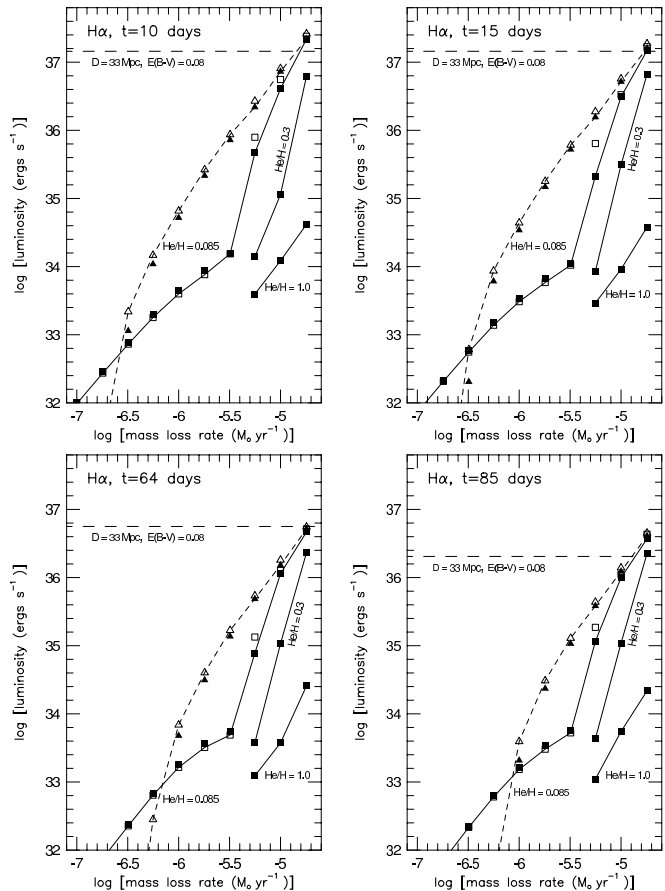}}
\caption{
H$\alpha$ luminosity as a function of mass loss rate at four epochs
(after the explosion) corresponding to the epochs listed in Table 2 for 
SN 2000cx, assuming a wind speed of $10 \kms$. The lines and symbols have the 
same meaning as in Fig. 3. The observed limits are for the four epochs in 
Table 2, assuming a wind temperature of $2.8\EE4$ K and a wind speed 
of $10 \kms$.
}
\label{fig12_letal}
\end{figure}

Allowing for a photospheric contribution to the ionization of the circumstellar gas,
the situation could change significantly. The main change is that photospheric
emission is independent of the wind density, and that models with low
values of $\Mdot/v_{w}$ become significantly more ionized. For example, for
a model with $\Mdot/v_{w} =  3.2\EE{-6} v_{w10}^{-1} \Msunyr$ 
and $T_{e} = T_{rev} /2$, $R_{HII} / R_{s} \sim 1.5$ without the photospheric
contribution and $\sim 9$ with the photospheric contribution on day 10.
For $\Mdot/v_{w} =  1.0\EE{-5} v_{w10}^{-1} \Msunyr$ the corresponding
numbers are $\sim 6$ (cf. above) and $\sim 8$, respectively, i.e., the
role of the photospheric contribution decreases with increasing wind
density, as expected.

\subsubsection{Line emission}
Like in M05 for SN~2001el, we do not detect any narrow emission lines 
from the SN CSM in SN~2000cx (see below). In Figs. 3 and 4 we compare 
our derived luminosity limits for SN~2000cx to our model calculations to derive 
upper limits for the mass loss rate. Figure 3 shows results for all lines we have
considered for the earliest epoch, and Fig. 4 concentrates on
H$\alpha$ at four epochs. In all figures we have included models with
(triangles) and without (squares) a photospheric contribution to the 
ionizing radiation. Open symbols are for $T_{e} = T_{rev}/2$ models, and
filled symbols for $T_{e} = T_{i}$ models. Dashed lines join the most
likely (in terms of $T_{e}/T_{i}$, as estimated from the equipartition time for Coulomb
interactions) photospheric contribution models, and 
solid lines the most likely ``pure-reverse-shock-contribution" models.
Models have been run for three sets of He/H abundance (by numbers), namely
0.085, 0.3 and 1.0. 
Figure 4 shows how the line luminosities decay with time, and that the
most promising emission lines to look for are H$\alpha$, as well as 
for a large He/H-ratio, He~I~$\lambda10\,830$. We return to Figs. 3 and 4
in Section 4.3.

\subsection{Late emission}
In order to test the effects of hydrogen in the SN Ia ejecta on 
late time emission we have modeled late spectra using the code described in 
Kozma \& Fransson (1998), and Kozma et. al (2005), and also used for
SN 2001el in M05. The model results presented in M05
were further used by Leonard (2007) for SNe 2005am and 2005cf, and recently
by Shappee et al. (2013b) for SN~2011fe. Our calculations are based on the 
explosion model W7 (Nomoto et al., 1984; Thielemann et al. 1986), where 
we have artificially included varying amounts of solar abundance material.
We did four calculations in which we filled the central regions out 
to $10^3 \kms$ with 0.01, 0.05, 0.10, and 0.50 $\Msun$ of solar 
abundance material, respectively. No microscopic mixing is done in any of the 
models, and there is no macroscopic mixing of the solar abundance 
zones either. The maximum velocity of the solar abundance gas was fixed at $10^3 \kms$.
Mixing of solar abundance gas out to larger velocities, and/or other zones into the core, could 
affect the line fluxes somewhat, but the effects are likely to be small. For larger velocities of the 
hydrogen-rich gas, the most important effect would be to depress the height of the line profiles 
so that more hydrogen-rich gas can be accommodated when compared to observations. However, 
in none of the models of Marietta et al. (2000), is the characteristic velocity for the ablated gas in 
excess of  $10^3 \kms$.

The resulting spectra at two epochs, 270 and 380 days after explosion,
are shown in Fig.~5. The upper spectrum in each of the four panels shows the results 
from the model containing $0.50 \Msun$ of solar abundance material. While the overall 
spectra are discussed in Kozma et al. (2005), we concentrate here on the 
narrower $\sim 10^3 \kms$ lines. In Fig. 5, we clearly see a number of such 
narrow features originating from the central, hydrogen-rich region, i.e., 
[O~I] $\wll 6300, 6364$, $\Ha$ and ${\rm Pa}\alpha$, [Ca~II] $\wll 7291, 7324$ 
and the IR-triplet. The second and third spectra in each panel, from 
models containing 0.10 and 0.05 $\Msun$ of solar abundance material, show
decreasing strengths of these features. In the lowest spectra, from the model 
containing only $0.01 \Msun$ of solar abundance material, the features are no
longer distinguishable in these plots.

\begin{figure}
\center{\includegraphics[width=87mm, clip,]{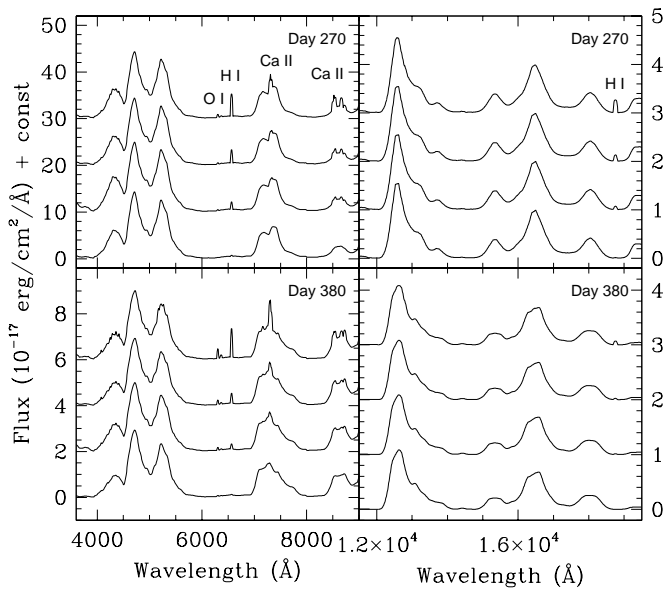}}
\caption{Modeled spectra at two epochs, 270 and 380 days after the
$B$-band maximum, for four models, with varying mass of the central, hydrogen 
rich region. The hydrogen mass increases from bottom to top in each frame, 
i.e., the lowest spectrum in each frame shows the model containg 0.01$\Msun$
of solar abundance matter, while the following spectra
contain 0.05, 0.10, and 0.50 $\Msun$, respectively.
In these spectra we have also labeled the features originating
in the hydrogen rich matter. These models were made for the distance and extinction
for SN~2001el (M05), i.e., $D = 17.9$ Mpc and $A_V =0.78$. See text for more details.
}
\label{fig13_letal}
\end{figure}

\begin{table*}
\caption{Log of VLT/UVES observations of SN 2000cx}
\begin{tabular}{lcccc}
\hline\hline
 UT date & Dichroic & Cross Disp/Filter/ & Wav. coverage & Exposure \\
         &          & /Central Wav.(nm)  & (\AA)         & (s)      \\
\hline
2000 July 20.3 & DIC1 & CD2/CUS04/390  &  $3300-4515$               &  $2\times2400$  \\
               &      & CD3/SHP700/564 &  $4620-5595$, $5675-6645$  &  \\
               & DIC2 & CD2/CUS04/437  &  $3755-4975$               &  $2\times2400$  \\
               &      & CD4/OG590/860  &  $6710-8525$, $8665-10300$ &  \\
2000 July 25.3 & DIC1 & CD2/CUS04/390  &  $3300-4515$               &  $3\times3600$  \\
               &      & CD3/SHP700/564 &  $4620-5595$, $5675-6645$  &  \\
2000 Sept 12.3 & RED  & CD3/SHP700/580 &  $4760-6840$               &  $1\times5400$  \\
2000 Oct 3.2   & RED  & CD3/SHP700/580 &  $4760-6840$               &  $2\times5400$  \\
\hline
\end{tabular} \\
\end{table*}

\section{Observations and Results}

\subsection{SN 2000cx}

\subsubsection{UVES observations and data reductions}
Our first VLT/UVES (D'Odorico \& Kaper 2000)
observation of SN 2000cx was obtained on 2000 July 20.3 UT,
already 6.4 days before maximum.
The supernova was observed using two dichroic standard
settings (DIC1 390+564, DIC2 437+860), enabling simultaneous observations
in the red and the blue. This effectively covers the full region of the
optical spectrum ($3260-10,600$~\AA). We used $2\times2400$s per setup. The 
night was clear and the seeing during the observations was typically 
less than 0\farcs7. All observations used the 0\farcs8 slit. 

A second epoch of observations was obtained 
on July 25.3 UT, i.e., 5 days after the first epoch, but still 1.4 days
before maximum. On this night only one mode was used (DIC1 390+564) covering
the wavelenghts $3260-4450$~\AA\ and $4580-6680$~\AA, and three integrations of
3600s each were obtained. Like the first night, the second night was also
clear, with a seeing below 0\farcs7. Again, the slit width 
was 0\farcs8. Details of the observations are compiled in Table 1. 

We also included other high resolution spectroscopic observations of SN 2000cx
made with VLT/UVES between 2000 September 12 and October 3 (48--69 days post
maximum) (cf. Patat et al. 2007a). For these observations the red arm standard
setting 580 had been used giving a wavelength coverage between 4760 and
6840~\AA. The slit width used was 1\farcs0. We selected
data from both September 12.3 and October 3.2 to be used in this study.

All the 2D data were interactively reduced using the UVES-pipeline
\footnotemark \footnotetext{http://www.eso.org/observing/dfo/quality/(version 2.0)} as
implemented in $\tt MIDAS$ and calibration frames obtained during daytime
operation. This included creating master bias and flat field frames, bias
subtraction, and flat fielding the data in the pixel space, and the geometric
calibration for the data. The 2D data were background subtracted and extracted
using standard IRAF tasks and a telluric correction was performed using
spectra of standard stars observed just before or after the SN spectra.
We describe these procedures in detail in M05.

\begin{figure}
\center{\includegraphics[width=85mm, clip, angle=0]{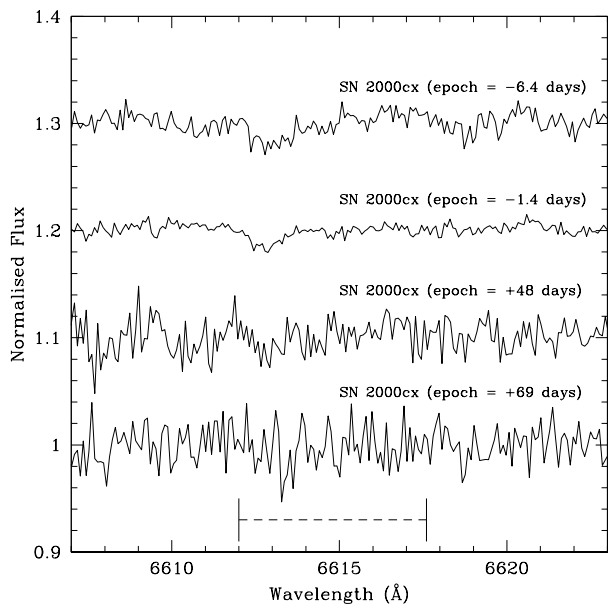}}
\caption{
Normalized UVES spectra in the expected spectral region around H$\alpha$
for SN 2000cx on four epochs (relative to $B$-band maximum) July 20.3, July 
25.3, September 12.3, and October 3.2 (UT) 2000. The expected wavelength range 
of H$\alpha$ is marked with a horizontal dashed line. No signs of circumstellar H$\alpha$ lines are
visible either in emission or absorption. Note that the feature at 6613~\AA\
visible in the two earliest spectra is likely due to a diffuse interstellar band (cf. Sect. 3.1.1). 
The $-$6.4, $-$1.4, and +48 day spectra have been shifted vertically for clarity.
}
\label{fig1_letal}
\end{figure}
We note that UVES flat fields contain telluric features that could lead
to spurious residual spectral features in the reduced science spectra. This is
due to the long optical path length inside the spectrograph. In the region
near H$\alpha$ at $\sim$6622~\AA\ and $\sim$6626~\AA\ there are features
present in the two first epoch (high S/N) spectra that were not fully
removed by the flat-fielding. However, we found that these residual
features were removed very well when dividing by the standard star
spectrum observed just after or before the science frames. Another weak
($\sim$ 2\% level) absorption feature (FWHM~$\sim$1~\AA) is also present in
our high S/N data at $\sim$6613~\AA. We found no signs of this feature
in either the dome flat fields or the standard star spectra (with similar
S/N to the SN spectra) observed just before or after the SN. Therefore, it is
unlikely that this feature could be due to any instrumental or atmospheric
effects. However, the same feature has also been seen in some other high S/N
UVES observations, e.g., HD 208669 observed on 2000 June 16 (Burkhard Wolff,
private communication) indicating a likely Galactic origin. 
It could be due to the known 6613.56~\AA\ diffuse interstellar
band (DIB) which has an intrinsic width slightly broader than 1~\AA.
From the central depth of $\sim 2$\% and the FWHM of $\sim 1$~\AA, we obtain an 
estimated equivalent width of $\sim 20$~m\AA. If we use Table~3 of Luna et al. (2008),
which lists an equivalent width of 210~m\AA\ in 6613.56~\AA\ for each unit of galactic 
$E(B-V)$, our estimated equivalent width indicates a galactic extinction 
of $E(B-V) \sim 0.1$ toward SN~2000cx.

\begin{figure*}
\center{\includegraphics[width=160mm, clip, angle=0]{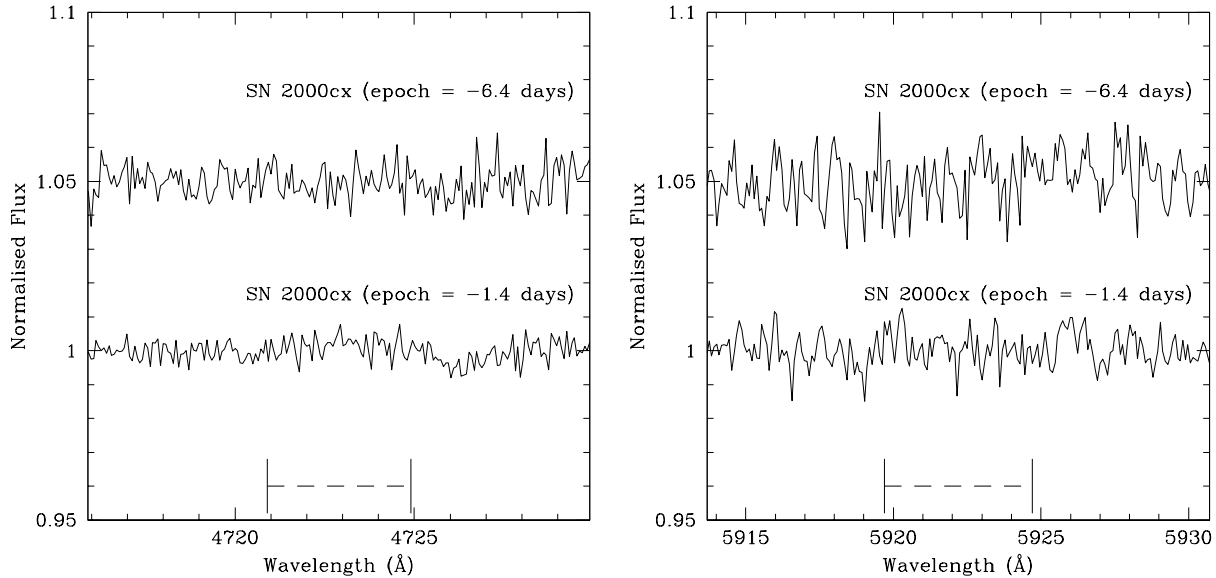}}
\caption{
Normalized UVES spectra in the expected spectral regions around 
He~II~$\lambda4686$ and He~I~$\lambda5876$ for SN 2000cx on two epochs
(relative to $B$-band maximum) July 20.3, July 25.3 (UT) 2000. The expected 
wavelength range of the circumstellar line is marked with a horizontal dashed line. No 
signs of circumstellar lines are visible either in emission or absorption.
}
\label{fig2_letal}
\end{figure*}

Flux calibration was achieved by comparison to low-resolution spectra
(Li et al. 2001) observed at similar epoch to the UVES observations
July 23, July 26, September 6, and October 6. These spectra were in turn
re-calibrated in absolute flux to match the observed $V$-band photometry
(Li et al. 2001) at the dates of our UVES observations. The fluxing is expected
to be accurate to better than $\pm$20$\%$.

\subsubsection{Circumstellar lines}
The host galaxy of SN 2000cx, NGC 524, has a NED recession velocity of
$2379 \pm 15 \kms$, consistent with the recession velocity estimate of
Emsellem et al. (2004) of $2353 \kms$, and and has observed maximum rotation 
velocity of $124 \kms$ (Simien $\&$ Prugniel 2000). We have therefore sought for 
any sign of narrow Balmer lines, and He I ($\lambda 5876$) and He II ($\lambda 4686$) 
lines at the expected wavelenghts $\pm125\kms$. No signs of such CSM lines 
are visible neither in emission, nor in absorption in our data. Figures 6 and 
7, respectively, show the region of $\Ha$ for all the four epochs of
observation, and the regions around the helium lines for the two first
epochs of observation.

\begin{table}
\caption{3$\sigma$ upper limits on circumstellar emission line fluxes of SN~2000cx. (Fluxes are not dereddened.)}
\begin{tabular}{lcccc}
\hline\hline
Julian day & Epoch$^{a}$ & Line & FWHM   & Flux \\
(2450000+)   & (days)      &      & (km~s$^{-1}$) & (ergs~s$^{-1}$~cm$^{-2}$) \\
\hline
1745.8  &$-6.4$ & H$\alpha$          & 37$^{b}$ & 8.4($-17$)$^{c}$ \\
        &       & H$\alpha$          & 62$^{d}$ & 1.3($-16$) \\
        &       & H$\alpha$          &106$^{e}$ & 1.5($-16$) \\
        &       & H$\alpha$          &203$^{f}$ & 2.3($-16$) \\
        &       & He~I$^{g}$         & 21$^{b}$ & 7.2($-17$) \\
        &       & He~I$^{g}$         & 53$^{d}$ & 1.1($-16$) \\
        &       & He~II$^{h}$        & 21$^{b}$ & 4.7($-17$) \\
        &       & He~II$^{h}$        & 53$^{d}$ & 7.9($-17$) \\
1750.8  &$-1.4$ & H$\alpha$          & 37$^{b}$ & 8.7($-17$) \\
        &       & H$\alpha$          & 62$^{d}$ & 1.4($-16$) \\
        &       & He~I$^{g}$         & 21$^{b}$ & 7.2($-17$) \\
        &       & He~I$^{g}$         & 53$^{d}$ & 1.3($-16$) \\
        &       & He~II$^{h}$        & 21$^{b}$ & 4.0($-17$) \\
        &       & He~II$^{h}$        & 53$^{d}$ & 7.7($-17$) \\
1799.7  & +47.5 & H$\alpha$          & 37$^{b}$ & 3.1($-17$) \\
        &       & H$\alpha$          & 62$^{d}$ & 4.8($-17$) \\
1820.7  & +68.5 & H$\alpha$          & 37$^{b}$ & 1.3($-17$) \\
        &       & H$\alpha$          & 62$^{d}$ & 1.7($-17$) \\
        &       & H$\alpha$          &106$^{e}$ & 2.0($-17$) \\
        &       & H$\alpha$          &203$^{f}$ & 3.0($-17$) \\
\hline
\end{tabular} \\
$^{a}$Relative to $B$-band maximum (JD2451752.2, Li et al. 2001). To obtain
the time since explosion used in, e.g., Figs. 11 and 12, a rise time of
16 days was assumed (as found for SN 1994D in Riess et al. 1999).\\
$^{b}$Assuming $T=2.8\EE4$~K and $v=10\kms$ for the wind.\\
$^{c}$9.2($-17$) stands for $9.2\EE{-17}$.\\
$^{d}$Assuming $T=2.8\EE4$~K and $v=50\kms$ for the wind.\\
$^{e}$Assuming $T=2.8\EE4$~K and $v=100\kms$ for the wind.\\
$^{f}$Assuming $T=2.8\EE4$~K and $v=200\kms$ for the wind.\\
$^{g}$He~I~$\lambda$5876 \\
$^{h}$He~II~$\lambda$4686
\end{table}

To estimate $3\sigma$ detection limits for narrow CSM emission or
absorption lines, we rebinned the spectra to have a Nyquist sampling,
i.e., one pixel equal to FWHM/2 for each expected line width (see Table 2).
The pixel-to-pixel standard deviations were then measured from the
rebinned spectra in a $\pm$3 $\times$ $125\kms$ region around the most
probable location of each CSM line. The reason for measuring the noise
in a larger wavelength region with similar noise characteristics to the
actual $\pm125\kms$ CSM line search region was to obtain better statistics
in the rebinned spectra. The $1\sigma$ flux levels were then determined by 
measuring the flux of Gaussian profiles with the relevant FWHM and their peaks 
equal to the measured pixel-to-pixel standard deviations. In our earlier
study (M05) this simple approach was found to give consistent 
upper limits for similar UVES data with a more mathematically robust method 
based on investigation of the noise histograms. The derived 3$\sigma$ upper 
limits for the hydrogen and helium lines are listed in Table 2 for H$\alpha$ 
at four epochs and the helium lines at two epochs. For the first and fourth 
epoch of observation H$\alpha$ upper limits are also given for wind velocities 
of 100 km~s$^{-1}$ and 200 km~s$^{-1}$. These limits are based on noise
measurements in $\pm$8 $\times$ $125\kms$ wavelength regions around the
expected location for the CSM $\Ha$. Such higher wind velocities could be 
expected if the wind would be radiatively accelerated by the supernova. Acceleration 
to these velocities will compress the gas ahead of the blast wave, but the overall density
profile will not deviate more than slightly from $\rho \propto r^{-2}$ after a few days, and 
our analysis will still be correct, using the unaccelerated wind velocity in estimates 
of $\Mdot/v_{w}$.

In Sect. 4, the corresponding line fluxes are compared to the models,
and upper limits for the mass loss rates are derived for the SN 2000cx 
progenitor system. The limits depend on the macroscopic velocity and expected 
thermal broadening of the lines. Table 2 gives 3$\sigma$ limits for different wind
velocities but only for one single temperature $2.8\EE4$~K which is a typical 
temperature found for the H$\alpha$ emitting gas (cf. Sect. 2.1.2).

\begin{figure}
\center{\includegraphics[width=60mm, clip, angle=-90]{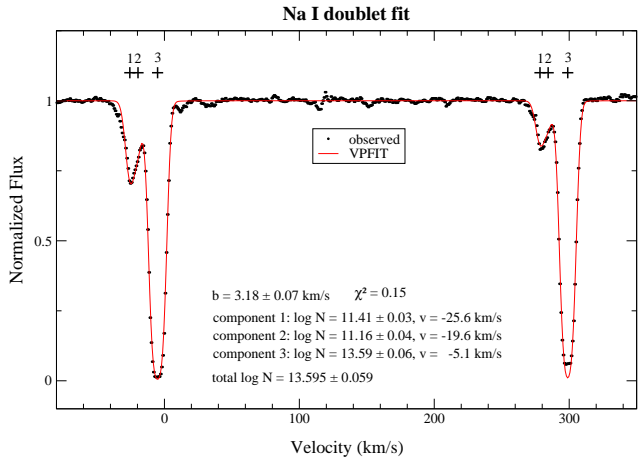}}
\caption{Na~I~$\lambda\lambda$5890, 5896 absorption in the Galaxy as seen
toward SN 2000cx. The velocity scale is heliocentric, and centered on 
the $\lambda$5890 component. The galactic absorption 
is dominated by a component at $\sim -5 \kms$. There is also weaker blueshifted 
absorption around $20-25 \kms$, and the absorption can be traced on the
blue side to $\sim -40 \kms$. Our VPFIT models disentangle three 
components. Their velocities and corresponding Na~I column densities are
shown in the figure, as well as in Table~3.
}
\label{fig3_letal}
\end{figure}

\begin{figure}
\center{\includegraphics[width=70mm, clip, angle=-90]{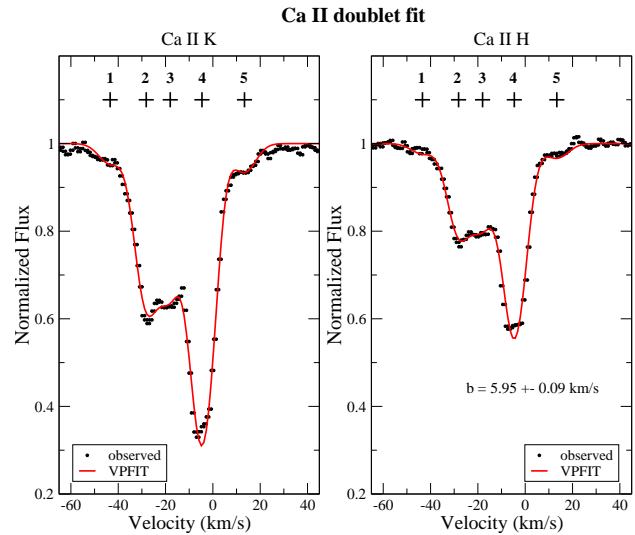}}
\caption{Ca~II~$\lambda\lambda$3933, 3968 absorption in the Galaxy as seen
toward SN 2000cx. The velocity scale is heliocentric. Note that the galactic 
absorption is dominated by the same component around $-5 \kms$ as in Fig. 3, 
but that there are two more components than for Na~I. For Ca~II we identify the 
following components: $-43.4\pm0.6$, $-28.1\pm0.3$, $-18.0\pm0.3$, 
$-4.6\pm0.1$ and $13.4\pm0.3 \kms$. The column densities for these components 
are given in Table~3.
Components `2', `3' and `4' (as marked in the figure) agree in velocity 
with the three Na~I components in Fig.~8.
}
\label{fig4_letal}
\end{figure}

\begin{figure}
\center{\includegraphics[width=70mm, clip]{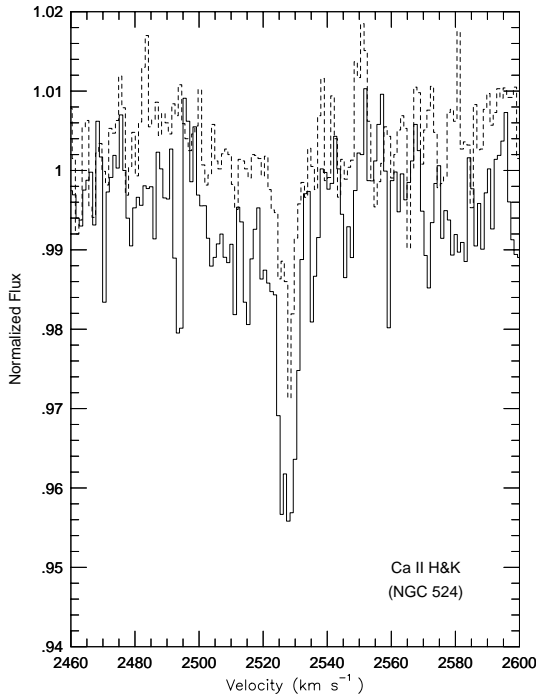}}
\caption{Weak Ca~II H ($\lambda $3968, dashed) and K ($\lambda $3934, solid)
absorption can also be seen from the host galaxy of 
SN 2000cx with maximum absorption at $2526.4\pm0.3 \kms$, which is $147 \kms$ 
higher than the NED value for NGC 524. The Ca~II column density in the host galaxy 
is $(5.5\pm0.4)\EE{10} \cm2$. No absorption due to Na I is seen from NGC 524. The velocity 
scale is heliocentric.}
\label{fig5_letal}
\end{figure}

\begin{table}
\caption{Galactic interstellar absorption components toward SN~2000cx.}
\begin{tabular}{lcc}
\hline\hline
Line & velocity$^{a}$ &  log(column density) \\
     & (km~s$^{-1}$)       &     (cm$^{-2}$) \\
\hline
Na~I~D      & $-25.6\pm0.3$ & $11.41\pm0.03$ \\
            & $-19.6\pm0.3$ & $11.16\pm0.04$ \\
            & $-5.1\pm0.1$ &  $13.59\pm0.06$ \\
\\
Ca~II~H\&K  & $-43.4\pm0.6$ & $10.93\pm0.04$  \\
            & $-28.1\pm0.3$ & $11.91\pm0.01$  \\
            & $-18.0\pm0.3$ & $11.85\pm0.01$  \\
            & $-4.6\pm0.1$  & $12.33\pm0.01$  \\
            & $+13.4\pm0.3$ & $11.09\pm0.03$  \\
\hline
\end{tabular} \\
$^{a}$Heliocentric velocity.\\
\end{table}

\subsubsection{Interstellar lines and reddening}
The estimates in Table 2 did not take into account any reddening. The 
galactic reddening in the direction of NGC 524
is $E(B-V)=0.083$ (Schlegel et al. 1998). Burstein \& Heiles (1982) found
a lower value of $E(B-V)=0.03$ based on an H~I/galaxy counts method. We have 
also estimated the amount of reddening from the Na~I~D absorption lines in 
our data. We have done this for epoch 2 when the supernova was brighter. 
The galactic Na~I~D lines at 5889.95~\AA\ and 5895.92~\AA\ have EWs of
0.36~\AA\ and 0.29~\AA, respectively, which is similar to what Patat et al. (2007a)
find.  As shown in Fig.~8, the lines seem to be made up of two major components, 
with the most redshifted being the strongest. No Na~I~D absorption at the velocity of 
NGC 524 could be detected. The pixel-to-pixel noise indicates an upper limit of an 
unresolved line of 4.2 m\AA, which shows that the extinction in the supernova 
host galaxy is indeed very low.  Patat et al. (2007a) estimate an EW of $\lsim 1$m\AA.

Also galactic absorptions in Ca~II~H and K are clearly present. 
The EW of Ca~II~K (3933~\AA) is 0.23~\AA\ and of Ca~II~H (3968~\AA) 0.14~\AA,
again in perfect agreement with Patat et al. (2007a). The Ca~II~H absorption 
line profile (from our 25 July spectrum) is shown in Fig.~9. It is clear that the same red 
component at $-5 \kms$ as for Na~I~D also dominates Ca~II, but that 
the bluer components ($-18 \kms$ and $-28 \kms$) are stronger and bracket 
the blue component of Na~I~D (cf. Fig. 8). However, if we allow the blue Na~I
component to be split up in two subfeatures using the program
VPFIT\footnotemark \footnotetext{http://www.ast.cam.ac.uk/$\sim$rfc/vpfit.html} 
(see Sollerman et al. 2005 for our use of VPFIT) as indicated in Fig. 8, their velocities 
coincide rather well with components of Ca~II. Unlike Na~I, Ca~II has also absorption 
components at $\sim -43 \kms$ and $\sim +16 \kms$.
The galactic components we identify using VPFIT, and their corresponding
column densities are listed in Table~3 for both Na~I and Ca~II. The VPFIT
models were made to fit both components of Na~I simultaneously, and likewise
the Ca~II H \& K components simultaneously. The different velocity components
for each ion were assumed to have the same FWHM (the parameter $b$ in Figs. 8
and 9), which should be close to the instrumental resolution. Patat et al. (2007a) 
argue for a split of our $-4.6\pm0.1 \kms$ component into two components, and 
an extra component at $-23.1\pm0.1 \kms$, but do not identify our component 
at $-43.4\pm0.6 \kms$.

\begin{figure*}
\center{\includegraphics[width=100mm, clip, angle=90]{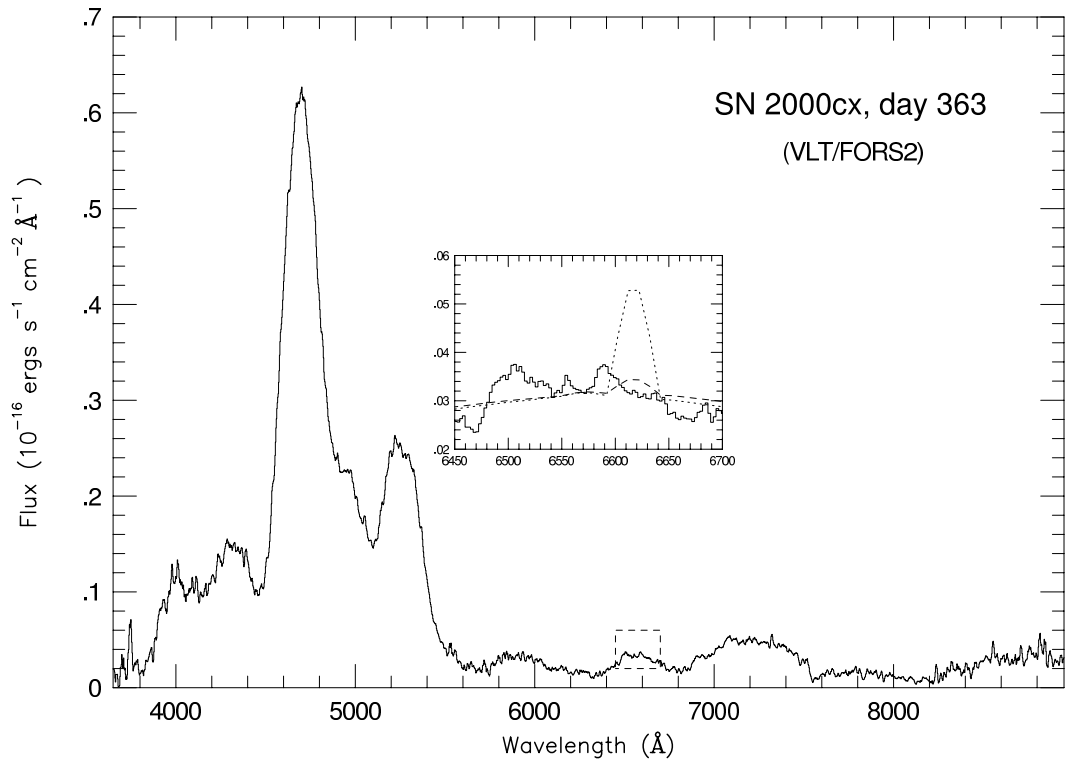}}
\caption{VLT/FORS2 spectrum of SN 2000cx at 363 days after $B$ maximum.
The overall spectrum is described in Sollerman et al. (2004). The inset shows a 
detailed view of the region marked by dashed lines, concentrating on wavelengths
around H$\alpha$, i.e., $\sim 6615$~\AA\ at the redshift of the supernova.
The dashed and dotted spectra in the inset are models for the supernova,
assuming 0.01 and 0.05$\Msun$ of solar abundance material, respectively,
concentrated to $\pm 10^3 \kms$ around the recession velocity of the supernova.
This is to simulate H$\alpha$ emission from gas removed from a possible
companion star according to the models of Marietta et al. (2000). Our models
for the H$\alpha$ emission are described in the text. There is no sign of
H$\alpha$ emission in SN 2000cx at these wavelengths, which places a limit
on hydrogen-rich material at velocities below $10^3 \kms$ of $\sim 0.03 \Msun$.
}
\label{fig6_letal}
\end{figure*}

\begin{figure*}
\center{\includegraphics[width=100mm, clip, angle=90]{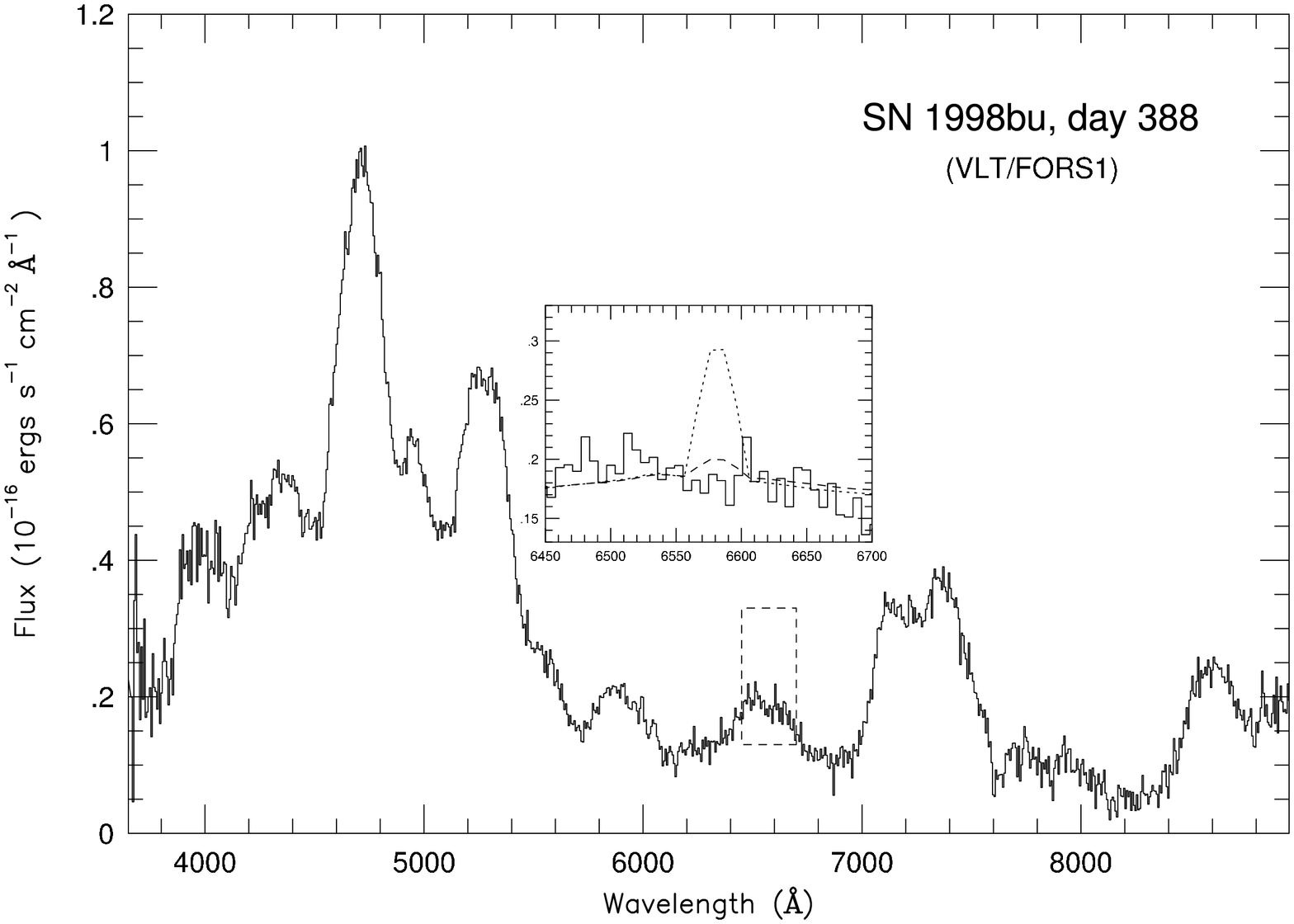}}
\caption{VLT/FORS1 spectrum of SN 1998bu at 388 days after $B$-band maximum.
The overall spectrum is described in Spyromilio et al. (2004).
The inset is a blow-up of the region marked by dashed lines, concentrating on 
wavelengths around H$\alpha$. As in Fig. 11 we have plotted our model for 
H$\alpha$ emission from a possible ablated companion star, and the mass within
the core (i.e., $\pm 10^3 \kms$) is the same as in Fig. 11. We have assumed
a distance to SN~1998bu of 11.2 Mpc, an extinction of A$_V = 1.0$ and a
redshift of $750\kms$ (Munari et al. 1998; Centurion et al. 1998).
There is no sign of H$\alpha$ emission in SN 1998bu at the expected wavelength.}
\label{fig7_letal}
\end{figure*}

Although much fainter, 
the Ca~II~H and K lines are also detected in NGC 524 (Fig. 10). The EW of the 
host galaxy lines are about 2.9 and 1.5m\AA, respectively, i.e., roughly 
100 times weaker than the Galactic lines. The redshift of these lines 
correponds to a velocity of 2526 km~s$^{-1}$, which is 147 (or 173) km~s$^{-1}$ higher 
than the recession velocity of NGC 524 according to NED (or Emsellem et al. 2004). 
This may be due to the rotation of NGC 524 at the position of the absorbing gas. 
Using VPFIT, the Ca~II column density in NGC 524 toward the supernova is a 
mere $(5.5\pm0.4)\EE{10} \cm2$. This low value is not surprising given the 
galaxy type of NGC 524 (S0). Patat et al. (2007a) find a column density 
of $(4.4\pm0.5)\EE{10} \cm2$, i.e., consistent with our value, as well as 
only $3 \kms$ difference in recession velocity.

We have used our EWs in combination with the expressions of Munari \& Zwitter 
(1997) to estimate the amount of Galactic reddening in the direction of the 
supernova. According to their Table 2, an EW of 0.36~\AA\ in Na~I correponds 
to $E(B-V) \simeq 0.15$, i.e., higher than the NED value given by 
Schlegel at al. (1998). However, as illustrated in Figs. 8 and 9, and in 
Table 3, the interstellar absorption toward SN~2000cx is composed of at least
three components which means that there is no simple relation between $E(B-V)$ 
and Na~I column density. In particular, for a saturated line like the central 
component of the Na~I~D lines in our data, the EW scales 
roughly as $\propto$~lg($N_{\rm col}$), whereas adding unsaturated components 
outside the core of the saturated one makes the EW scale with the total column 
density as $\propto N_{\rm col}$. Added to this uncertainty due to the 
cloud distribution, the EW vs. $E(B-V)$ relation is also sensitive to the 
degree of ionization along the line of sight to the supernova, as well as to 
the gas-to-dust ratio. With these caveats in mind, Na~I~D absorption can 
therefore, at best, only provide a rough estimate of the extinction. We have
discussed this further for SN~2001el in Sollerman et al. (2005).

A further constraint on the extinction of SN 2000cx comes from the supernova
itself. In Li et al. (2001), the late time color of SN 2000cx is shown
to be very blue. When corrected for a Galactic extinction of $E(B-V)=0.15$,
it is indeed bluer than for most known unreddened SNe~Ia. This argues that 
the extinction is low toward SN 2000cx, and that the Na~I~D absorption 
could give too high an estimate of $E(B-V)$ using the simple Munari \& Zwitter 
approach. As mentioned in Sect. 3.1.1, the likely DIB feature at 6613.56~\AA\ 
provides an estimate of $E(B-V)$, and we found $E(B-V) \sim 0.1$, which would
support the Schlegel et al. galactic reddening estimator. It should be cautioned that the 
scatter on the EW/$E(B-V)$ relation for the DIB could easily be 50\%, and possibly even 
a factor 2, in particular for very low reddening sightlines. However, as there seems to be no
strong evidence against using the Schlegel et al. estimate of $E(B-V)=0.08$, we have in 
the following used this value. This yields a flux correction 
factor of 1.20 for H$\alpha$, 1.23 for He~I~$\lambda$5876, and 1.32 for
He~II~$\lambda$4686 for the limits obtained for the circumstellar emission 
in Sect. 2.1.2.

\subsubsection{FORS observations and data reductions}
On 2001 July 24 we obtained a 2400~s exposure using FORS2 on UT4. We used the 
300V grism together with order sorting filter GG375 and a $1\farcs3$ wide
slit. Two nights later, on 2001 July 26, this spectrum was complemented with
another 2400~s exposure with the 300I grism and the OG590
filter. Together these two spectra cover the wavelength region
$\sim 3700 - 9200$~\AA.

The spectra were reduced in a standard way, including bias
subtraction, flat fielding and wavelength calibration using exposures of
a Helium-Argon lamp.  Flux calibration was done relative to the
spectrophotometric standard star  LTT377 in 300V and to G158-100 in
300I. The absolute flux calibration of the combined spectrum was
also checked against broad band photometry of the SN.

\subsubsection{Late emission}
In Fig. 11 we show the combined optical spectrum (2001 July 24 and 26)
of SN 2000cx. The average epoch corresponds to day 363 after maximum. The 
general features of the spectrum are discussed in Sollerman et al. (2004) 
along with the late photometry of the supernova. Here we use 
the spectrum for an inspection of possible signatures of gas from a
tentative hydrogen-rich binary companion, as described in the models of 
Marietta et al. (2000). The inset shows a detailed view of the spectral region 
around H$\alpha$, along with our model calculations discussed in Sect. 2.2. 
We see no evidence of emission from gas originating from a companion.

\subsection{SN 1998bu}
\subsubsection{FORS observations and data reductions}
We have also made a similar test for our late spectra of SN~1998bu.
The supernova was observed using FORS1 on VLT on 1999 June 11
corresponding to 388 days past maximum. These observations and data
reductions are described in Spyromilio et al. (2004).

As shown by Cappellaro et al. (2001), SN 1998bu started to deviate from
the normal SN Ia decay rate some time after $\sim 300$ days, and by $\sim 500$
days it was brighter by $\sim 2$ magnitudes than normal SNe Ia (e.g., SN 1996X).
The reason for this is that SN 1998bu was strongly affected by a light echo. 
The magnitude we estimate on day 388 places SN 1998bu less than half a 
magnitude (in $V$-band) from the ``light echo free'' $V$-band light curve 
shown in Fig. 1 of Cappellaro et al. (2001). This means that the observation 
at 388 days is the latest phase available (see also Spyromilio et al. 2004) 
for SN 1998bu, while still rather unaffected by the light echo. 

\begin{figure}
\center{\includegraphics[width=85mm, clip,]{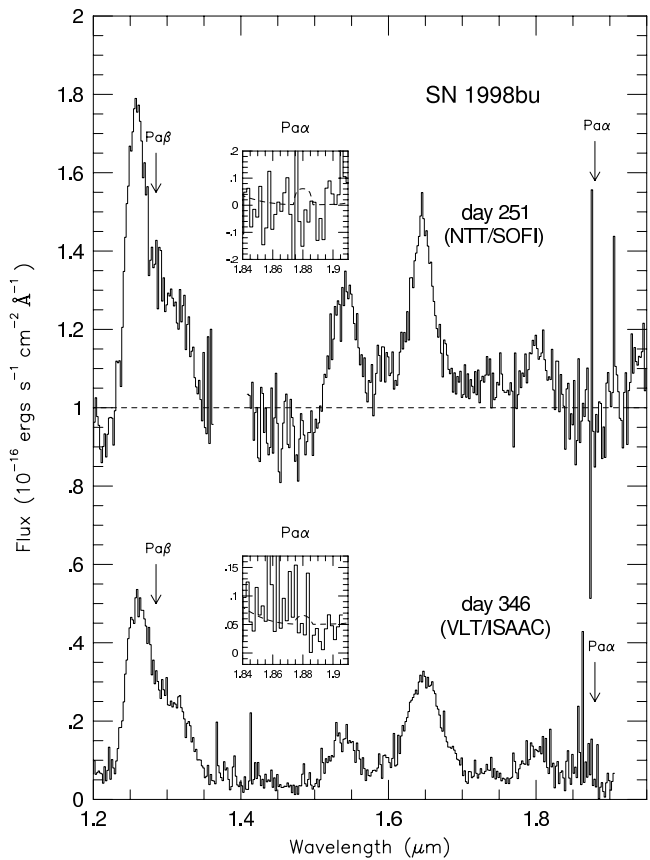}}
\caption{Infrared spectra of SN 1998bu in the nebular phase taken at 251
days (ESO/NTT/SOFI) and 346 days (ESO/VLT/ISAAC) after $B$ maximum. The
overall spectra are modeled and discussed in Spyromilio et al. (2004). 
The day 251 spectrum has been shifted upwards 
by $10^{-17}$ erg~s$^{-1}$~cm$^{-2}$\AA$^{-1}$ for clarity, and its zero-level
is marked by a dashed line. The arrows mark the rest wavelenghts of
Pa$\alpha$ and Pa$\beta$ for SN 1998bu. The insets show blow-ups of the 
spectral region around Pa$\alpha$. The dashed line shows a model
discussed in Sect. 2.2 and displayed in Fig. 13 assuming $0.5\Msun$ of 
hydrogen-rich material within the innermost $\pm 10^3 \kms$ of the
ejecta. This is to simulate the emission from the companion star in the 
models of Marietta et al. (2000). However, there is no sign of such 
Pa$\alpha$ emission in SN 1998bu. Note that Pa$\alpha$ sits in a region of 
very poor atmospheric transmission, and that Pa$\beta$ is a less efficient 
emitter and sits in the wing of a strong ejecta line (see Spyromilio et al. 
2004).  
The modeled emission in the inset of the lower spectrum was shifted upwards 
by $5\EE{-18}$ erg~s$^{-1}$~cm$^{-2}$\AA$^{-1}$ to overlap with the observed
spectrum. Same reddening, redshift and distance were assumed as in Fig. 12.
}
\label{fig8_letal}
\end{figure}

The resulting spectrum is shown in Fig. 12 (see also Fig. 4 of Spyromilio et
al. 2004), and it is indeed very similar to the nebular spectrum at 330 days
published by Cappellaro et al. (2001). A more careful inspection shows that 
the emission blueward of $\sim 4500$\AA\ is relatively stronger in our day 
388 spectrum, which is likely to be a sign of the light echo, as also 
highlighted in Fig. 4 of Spyromilio et al. (2004). Of more importance to the 
current analysis is the region around H$\alpha$ at the expected redshift of 
SN~1998bu. We have also included the results from the model calculations 
discussed in Sect. 2.2 for the H$\alpha$ emission expected in the models of 
Marietta et al. (2000) if hydrogen is evaporated off a companion star. We find no 
sign of H$\alpha$ emission within this interval.

\subsubsection{NTT/SOFI and VLT/ISAAC observations}
We have also observed SN 1998bu in the infrared 251 days and 346 days 
after $B$ maximum with NTT/SOFI and VLT/ISAAC, respectively. Both these 
spectra (shown in Fig. 13) are safely within the nebular phase, and should be 
unaffected by the light echo. While Spyromilio et al. (2004) describe details 
of the observations, and also model the supernova emission, we concentrate 
here on the $\pm 10^3 \kms$ velocity intervals around the expected wavelengths 
of Pa$\alpha$ and Pa$\beta$ for the same reason as we did for H$\alpha$ in 
Sect. 3.2.1. 

To obtain a better signal-to-noise we have binned the spectrum 
in 20\AA\ bins, corresponding to $\sim 320 \kms$ at Pa$\alpha$. As for
H$\alpha$ in Sect. 3.2.1, we see no evidence of material from a companion 
star. The test in the infrared is, however, more complicated since Pa$\alpha$ 
sits in a region of very poor atmospheric transmission, and Pa$\beta$ is 
expected to be a less efficient emitter and sits in the wing of a strong 
ejecta line (see Spyromilio et al. 2004) which may have intrinsic structures 
that could confuse the search for $\lsim 10^3 \kms$ Pa$\beta$ emission. 

\subsection{Early Ca~II absorption}
As discussed in Sect. 1 several SN~Ia show high-velocity (HV) absorption
features in the Ca~II IR and H\&K lines. For the particular case
of SN~2000cx, Thomas et al. (2004) identified in the +2 day spectrum of Li
et al. (2001) HV features in both the IR and H\&K lines. While the IR
triplet have several subfeatures, the H\&K feature is wide and flat. The
velocity of the absorptions reaches out to $\sim 2.4\EE4 \kms$ and $\sim
2.2\EE4 \kms$, if we refer to the bluest components of the multiplets,
i.e., to $\lambda 3934$ and $\lambda 8498$\AA, respectively, and adopt a
recession velocity of $v_{rec}$ = 2379 km~s$^{-1}$ for the supernova. Any
changes in these velocities with time (up to +7 days) are difficult to
disentangle from the data presented in Thomas et al. (2004), although
there may be a hint that at least the IR feature may recede marginally.

Thomas et al. (2004) modelled the HV features assuming both spherical and
nonspherical distributions of the absorbing material. Their best fit was
obtained assuming both spherical symmetry and an elongated feature along
the line of sight with a maximum optical depth at $2.24\EE4 \kms$ (cf.
their Fig.~9). Interestingly enough, Leonard et al. (2000) reported
continuum polarization of $\sim 0.5\%$ for this supernova at +2 days,
which could in fact signal an aspherical scattering surface.

We have used our data to extend the study of HV Ca~II absorption in SN
2000cx to $-6$ days. As for our data of SN 2001el presented in M05
we decided to normalize the continuum by a first order
polynomial fit around the profile of interest to better trace the line
wings. The normalized spectra are shown in Fig.~14 where the solid line is
for the IR triplet, and the dashed line for the H\&K lines. The velocities
are with respect to the bluest components of the multiplets as discussed
for the data of Thomas et al. (2004) above.

\begin{figure}
\center{\includegraphics[width=87mm, clip,]{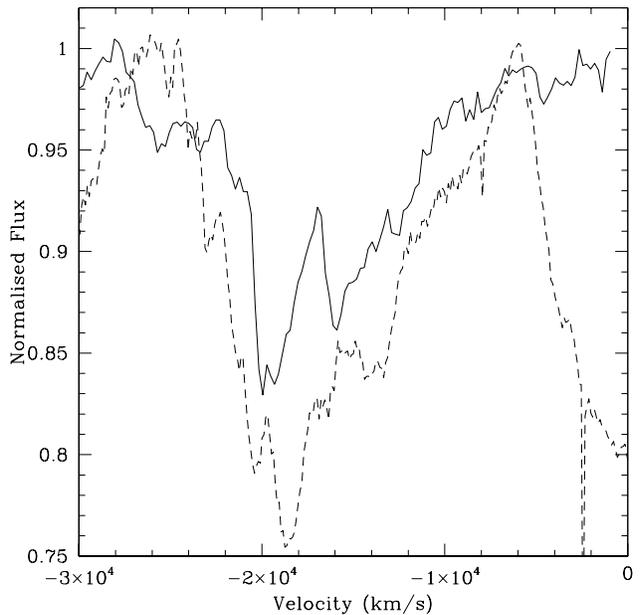}}
\caption{
Comparison between the Ca II IR triplet (solid line) and the Ca II H$\&$K
lines (dashed line) at $-6$ days. The flux calibrated ($f_{\lambda}$)
spectral features have been normalized and converted to velocity space
w.r.t. to the rest wavelengths assuming $v_{rec}$ = 2379 km~s$^{-1}$ for
SN 2000cx. For this we used the rest wavelengths of the bluest components
of the Ca II triplet/doublet lines viz. 3934~\AA\ and 8498~\AA.}
\label{fig14_letal}
\end{figure}

For the H\&K lines we find that the Ca~II absorption reaches out to $\sim
2.4\EE4 \kms$, and for IR lines at least out to $\sim 2.2\EE4 \kms$.
There is also an absorption feature in the IR lines reaching $\sim
2.8\EE4 \kms$ which, however, does not appear to be present in the H\&K
lines. Furthermore, it can be expected that the H\&K lines have a
higher optical depth (e.g., Mihalas 1978) and are present at a higher
velocity than the IR lines as they originate from a lower level of the
Ca~II ion and have larger oscillator strengths. In fact, the HV H\&K
lines have been observed to show wings extending to much higher velocities
than the IR lines at least in SNe 1990N and 2001el (for details see
M05). However, this behavior could also be
due to the H\&K profile blending with other lines (e.g., Kasen et al.
2003), and therefore a full NLTE analysis would be needed to be
conclusive for SN 2000cx. Also, the Ca~II line absorption shown
by SN 2000cx is much weaker than the one seen in SNe~1990N and 2001el,
making an estimate of the maximum velocity less accurate.

In contrast to the behavior shown by, e.g., SN 2001el (M05),
 it appears that in SN 2000cx the maximum Ca~II velocities at $-6$
days are similar to the maximum velocites at $+2$ days. In general, a
decreasing maximum velocity of the Ca~II absorption would be expected
simply because the column density of the ejecta decreases with time as
$\propto t^{-2}$, and disregarding abundance and ionization effects, the
most dilute gas will become optically thin faster. For the outer ejecta,
the density falls off quickly with radius (cf. Sect. 3.1) so a receding
blue edge of the IR triplet could fit into a picture where
the outermost ejecta become progressively optically thinner in Ca~II.
The constant maximum Ca~II velocity of SN 2000cx could indicate that 
the structure of the absorbing gas is more shell-like than in SN 2001el,
although we reiterate that conclusions from the weaker absorption in 
SN 2000cx must be drawn with caution.

\section{Discussion and Summary}
\subsection{Limit on wind density from absence of circumstellar line emission}
In Section 3.1.2 we derived upper limits for the H$\alpha$ line fluxes 
(see Table 2). Adopting an extinction of $E(B-V)$=0.08 and a distance of 33 Mpc
for SN2000cx, these flux limits correspond to line luminosities of
$L_{\rm H\alpha}$ = 1.3, 2.1, 2.4, and 3.7 $\times$ 10$^{37}$ erg~s$^{-1}$
for the first epoch observation and wind speed of 10, 50, 100, and 200
$\kms$, respectively. For the second and third epoch observations,
respectively, the upper limits for a 10 $\kms$ wind are 1.4 and 0.50 $\times$ 
10$^{37}$ erg~s$^{-1}$, and for a 50 $\kms$ wind they are 2.2 and 0.77 $\times$
10$^{37}$ erg~s$^{-1}$. For the fourth epoch observations our upper limits
correspond to $L_{\rm H\alpha}$ = 2.1, 2.7, 3.2, and 4.8 $\times$ 10$^{36}$ 
erg~s$^{-1}$ for the wind speeds of 10, 50, 100, and 200 $\kms$, respectively.
In these estimates we have assumed a luminosity-weighted temperature of
H$\alpha$ emitting gas of $2.8\EE4$ K. From Fig. 2 one can see that this
is a reasonable assumption since the temperature remains 
within $2.2\EE4 - 3.4\EE4$ K for all our $T_{e} = T_{i}$ models 
after $\sim 6$ days. The difference is small for models 
with $T_{e} = T_{rev}/2$. Table 2 also gives upper limits for 
He~I~$\lambda5876$ and He~II~$\lambda4686$ for the first two epochs using 
the same luminosity-weighted temperature. For the He~II line, this 
temperature is an underestimate, but as discussed below, this does not 
affect our conclusions.

Figure 3 shows that for the earliest epoch, the limit on $\Mdot/v_{w}$
from H$\alpha$ for a normal He/H abundance of the CSM 
and $v_{w10} = 1$ is $\sim 1.5\EE{-5} \Msunyr$. Should the wind be more 
helium-rich or faster, we cannot place any limit on $\Mdot$, neither from
H$\alpha$ nor He~I~$\lambda5876$ and He~II~$\lambda4686$, although 
the latter line comes close to the detection limit 
for $\Mdot \sim 2\EE{-5} v_{w10}^{-1} \Msunyr$ for $\rm{He/H} = 1$ 
and $v_{w10} = 1$. An even more promising helium-line to look for seems to be 
He~I~$\lambda10\,830$. This line could, however, not be covered by the UVES
observations.

Turning to the later epochs, we note no improvement in terms of the limit
on $\Mdot$ from H$\alpha$, except for 85 days post explosion, for which we find
the limit $\sim 1.3\EE{-5} \Msunyr$ in case of $\rm{He/H} = 0.085$ 
and $v_{w10} = 1$. For $v_{w10} = 5$ (and $\rm{He/H} = 0.085$), the limit 
goes up $\Mdot \sim 7.5\EE{-5} \Msunyr$, and for still higher $\Mdot$ the 
limit goes beyond $10^{-4} \Msunyr$. For $\rm{He/H} = 0.3$ and $v_{w10} = 1$
the limit is $\Mdot \sim 1.8\EE{-5} \Msunyr$. In the most favorable case 
of our models, the limit on $\Mdot$ is thus $\sim 1.3\EE{-5} \Msunyr$ (for 
a normal He/H-ratio as well as slow wind ($v_{w10} = 1$). A more helium-rich
wind degrades the limit somewhat, whereas a faster wind moves the 
limit on $\Mdot$ upwards quickly. The limit is also prone to systematic
uncertainties, in particular the assumptions for the ionizing spectrum, 
as well as the assumption of a smooth, spherically distribution of the CSM. 
How these effects affect the limit of $\Mdot$ was discussed in some 
detail in C96. 

\subsection{Comparing our circumstellar limits for SN~2000cx to other supernovae}
We have looked for narrow hydrogen and helium lines in our
early high-resolution optical spectra of the Type Ia supernova
SN~2000cx. No such lines were detected. We have made detailed 
time dependent photoionization models for the possible circumstellar
matter to interpret these non-detections. The best limit was obtained
from H$\alpha$, and we put a limit on the mass loss rate from the
progenitor system of $\sim 1.3\EE{-5} \Msunyr$ in case of $\rm{He/H} = 0.085$
and $v_{w10} = 1$, where $v_{w10}$ is the wind speed in $10 \kms$.
These limits are sensitive to gas outside a few $\times 10^{15}$~cm from the 
supernova. Possible circumstellar matter closer to the supernova would 
have been swept up before our observations.
The limit is slightly higher than that we found for SN~2001el (M05) 
for which we derived $\sim 9\EE{-6} \Msunyr$, but similar to
what C96 found for SN 1994D. However, the model for SN 1994D
assumed a much slower circumstellar shock and applying the model used
in this paper to the Cumming et al. study of SN 1994D would increase their
limit substantially.

The H$\alpha$ based upper limits on circumstellar matter around SNe Ia
described in M05 for SN~2001el and here for SN~2000cx are
the most sensitive so far. These results, together with the previous 
observational work on SN~1994D (C96)
disfavour a symbiotic star in the upper mass loss rate regime (so called
Mira type systems) from being the likely progenitor scenario for these SNe.
This is in accordance with the models of Hachisu et al. (1999a, 1999b),
which do not indicate a mass loss rate of the binary companion in the
symbiotic scenario higher than $\sim$10$^{-6}$ $\Msunyr$. Stronger mass loss
would initiate a powerful (although dilute) wind from the white dwarf that
would clear the surroundings of the white dwarf, and that may even strip off
some of the envelope of the companion. The white dwarf wind with its possible
stripping effect, in combination with the orbital motion of the stars, is
likely to create an asymmetric CSM. Effects of a more disk-like structure of
the denser parts of the CSM, and how this connects to our spherically symmetric
models, are discussed by C96 In general, lower values
of $\Mdot /V_{\rm wind}$ should be possible to trace in an asymmetric scenario,
although uncertainties due the inclination angle and the flatness of the
dense part of the CSM would also be introduced. However, to push the
upper limits down to the $\sim$ 10$^{-6}$ $\Msunyr$ level a much more nearby
($D \sim 5$ Mpc) SN Ia needs to be observed even earlier ($\sim 15$ days
before the maximum light) than our first epoch observation of SN~2001el
(M05). 

Similar, or even lower limits on $\Mdot /V_{\rm wind}$ can be obtained
from X-ray, and especially radio observations (e.g., Eck et al. 2002), but such 
observations cannot disentangle the elemental composition of
the gas in the same direct way as optical observations can do. 
The most constraining radio limit so far is that of $\sim$ 10$^{-9}$ $\Msunyr$ for the 
nearby SN~2011fe (Chomiuk et al. 2012) which essentially rules out all likely single
degenerate progenitor systems. One should, however, keep in mind that also
SN~2006X was radio silent, which was used by Chugai (2008) to argue for
a mass loss rate below $\sim$ 10$^{-8}$ $\Msunyr$ for the progenitor system
for that supernova, despite it possibly having circumstellar gas (Patat et al. 2007b). 
The main problem with X-rays and radio limits, as well as the method used in this paper,
is that they probe circumstellar gas through ongoing shock interaction. Circumstellar
gas in shell-like structures far away from the explosion site are better probed
by the absorption-line method (e.g., Patat et al. 2007b), though a disk-like
CSM may be missed if the line-of-sight is not in the disk plane. For nearby 
SNe~Ia it seems imperative that both high-resolution spectroscopy as early as possible
and up to the first $\sim 100$ days should be performed, as well as continued radio monitoring
for several years. SN~2011fe may provide a good testbed for such investigations.
Somewhat discouraging for the latter is that no SNe~Ia has been detected in radio,
despite many of them having been nearby, as well as observed over a range of
times since explosion (Eck et al. 2002). For example, SN~1937C and  
SN~1972A did not show any radio emission at ages 48.7 and $9.0-13.7$ years, 
respectively. It could be that circumstellar shells in some cases are so far out that
one has to wait until the remnant phase to see increased shock interaction
(Chiotellis et al. 2012). The obvious exceptions are members of the SN 2002ic like
supernovae. A particularly interesting case is PTF 11kx (Dilday et al. 2012) with
its circumstellar interaction turning on after maximum light. For this case, inverse Compton
scattering in the shocked gas should be less important than if interaction would
have started immediately after the explosion. Our models should therefore be
applicable to this supernova after the circumstellar shock had formed.

\subsection{Limit on hydrogen-rich gas from a companion}
In the insets of Figs. 11 and 12, models with $0.01 \Msun$ and $0.05 \Msun$ were 
compared with late time spectra of SNe~2000cx and 1998bu, respectively.
From these insets it is obvious that $0.05 \Msun$ of solar abundance material with a velocity 
of $\lsim 10^3 \kms$ should have been seen in both SN~1998bu and SN~2000cx.
Our calculations give a lower limit on the amount of solar abundance
material of $\sim 0.03 \Msun$ that would not escape detection in late  H$\alpha$ spectra 
of the two supernovae. A similar limit was obtained by M05 for the late emission of SN 2001el, 
and Leonard (2007) finds $\sim 0.01 \Msun$ for SNe~2005am and 2005cf using our models. As 
discussed in M05, the limits we obtain for the $\lsim 10^3 \kms$ solar abundance 
material in all those five supernovae should be conservative limits.
The limit on SN~2011fe by Shappee et al. (2013b) is even lower, $0.001 \Msun$, and rests on 
an extrapolation of our model simulations to masses below $0.01 \Msun$.

Figure 13 shows that the infrared spectrum of SN~1998bu would not reveal even $0.50 \Msun$ 
of hydrogen-rich gas at those wavelengths. Pa$\alpha$ (or Pa$\beta$) based limits on 
hydrogen-rich gas are therefore substantially higher than for H$\alpha$ (cf. Fig. 12), even with the
most powerful instruments presently available. However, future space missions such as the 
James Webb Space Telescope ({\it JWST}) could provide a powerful tool to detect 
Pa$\alpha$.

Comparing a $\sim$0.01-0.03 M$_{\odot}$ limit to the numerical simulations of 
Marietta et al. (2000) indicates that symbiotic systems with a subgiant, red giant or a 
main-sequence secondary star at a small binary separation are not likely progenitor scenarios
for these SNe. More recent modeling by Pakmor et al. (2008) shows that the material stripped
by the companion could be significantly lower than that estimated by Marietta et al. (2000). In one of the six models for main-sequence donors tested by Pakmor et al. (2008), the mass of stripped envelope was even lower than 
the $\lsim$0.01$_{\odot}$ limit of Leonard (2007). On the other hand, the most recent models (Pan et al. 2012) 
have again pushed the evaporated mass for hydrogen-rich donors up to levels surpassing our limits. 

A solution to any conflict between the ablated mass in impact models and our limits on hydrogen 
in the nebular phase could be twofold: either the separation between the companions is large, 
or the donor star is a helium-rich star.  (See Fig. 12 of Pan et al. 2012, which summarizes the yield of 
unbound masses in models made so far.) It is, however, not only the ablated mass that is important, 
but also the velocity distribution of this material. In our models we have distributed the material
evenly within a $1000 \kms$ sphere. A velocity much below $1000 \kms$ would push our limits down to 
below $\sim$0.01 M$_{\odot}$. Pan et al. (2012) shows that most of the mass lost has indeed
velocities below $1000 \kms$, peaking at $\sim 500-600 \kms$ in the case of a hydrogen-rich 
donor. Helium-rich companions, on the other hand, have velocities which peak in velocity 
around $1000 \kms$. In the helium-rich case, the ablated mass is in most models below 0.01 M$_{\odot}$, 
and the observational signature would obviously in this case not be hydrogen lines, but lines from 
other elements. Figure 5 shows that lines from O~I and Ca~II are promising probes.
There is also the possibility that all the five SNe~Ia with a maximum of $\sim$0.01-0.03 M$_{\odot}$ 
hydrogen-rich gas, like very likely SN~2011fe (Shappee et al. 2013b), 
stem from double degenerate systems. One issue that is not a matter of concern in our models is radiative 
transfer effects of any H$\alpha$ emission produced in the center of our SN~Ia models. In all cases, 
the H$\alpha$ emission emerges unattenuated, i.e., electron scattering 
and line transfer are unimportant. To make even more accurate predictions of possible 
hydrogen/helium line emission we should use nebular emission models with a larger line list to test this, 
in conjunction with hydro simulations like those of Pan et al. (2012). At the same time, further deep spectra of 
SNe~Ia in the nebular phase are needed to get better supernova statistics. Comparing with Fig. 12 of Pan et al.
 (2012), it seems hydrogen-rich donors with a separation of  $\lsim$5 times the radius of the donor may 
be ruled out for the in total five SNe discussed here, by M05 and by Leonard (2007). For SN~2011fe
one has very likely to resort to a double degenerate scenario.


\vskip 0.5cm
We thank Sergei Blinnikov and Elena Sorokina for discussions, and for making 
the models w7jz.ph and w7jzl155.ph available to us. This work was supported 
by the Swedish Research Council, the Swedish Space Board, and the Royal 
Swedish Academy of Sciences. We also thank the staff at Paranal for carrying out the 
service observations of SN~2000cx. PL acknowledges support the Wallenberg Foundation 
and the Kavli Institute of Theoretical Astrophysics where part of this work was done. SM 
acknowledges support from the EC Programme `The Physics of Type Ia SNe' 
(HPRN-CT-2002-00303; PI: W. Hillebrandt) and from the Participating Organisations of 
EURYI and the EC Sixth Framework Programe. JS acknowledges a Reserach Fellowship 
at the Royal Swedish Academy supported by a grant from the Wallenberg Foundation.  
 EB was supported in part by grants from DOE, NASA and the NSF. 


\begin{thebibliography}{}
\bibitem[Aldering (2006)]{Ald06} Aldering, G. 2006, ApJ, 650, 510 
\bibitem[Benetti et al.(2006)]{2006ApJ...653L.129B} Benetti, S., 
Cappellaro, E., Turatto, M., Taubenberger, S., Harutyunyan, A., 
\& Valenti, S.\ 2006, ApJL, 653, L129 
\bibitem[Benz et al. 1990]{b90} Benz, W., Cameron, A.~G.~W., Press, W.~H., \&
     Bowers, R.~L. 1990, ApJ, 348, 647
\bibitem[Blinnikov \& Sorokina 2000]{blso00} Blinnikov, S.~I., \& 
     Sorokina, E. 2000, A\&A, 356, L30
\bibitem[Blinnikov \& Sorokina 2001]{2001spuv.proc...84B} Blinnikov, 
     S.~I., \& Sorokina, E.~I. 2001, in Scientific prospects of the space 
     ultraviolet observatory SPECTRUM-UV, ed. B.~M. Shustov \& D.~S. Wiebe 
     (Moscow: GEOS) (in Russian), 84
\bibitem[Branch 1998]{bra98} Branch, D. 1998, ARA\&A, 36, 17
\bibitem[Branch et al.\ 1985]{bra85} Branch, D., Doggett, J.~B., 
     Nomoto, K., \& Thielemann, F.-K. 1985, ApJ, 294, 619
\bibitem[Branch et al.\ 1995]{bra95} Branch, D., Livio, M., Yungelson, L.~R.,
     Boffi, F.~R., \& Baron, E. 1995, PASP, 107, 1019
\bibitem[Branch et al.(2004)]{2004ApJ...606..413B} Branch, D. et al. 2004, ApJ, 606, 413
\bibitem[Burstein \& Heiles 1982]{bur82} Burstein, D., \& Heiles, C. 1982,
     AJ, 87, 1165
\bibitem[Cappellaro et al.\ 2001]{capp01} Cappellaro, E. et al., 2001, ApJ, 549, 215
\bibitem[Centurion et al.\ 1998]{cen98} Centurion, M., Walton, N., \&
     King, D. 1998, IAU Circ., 6918
\bibitem[Chandra et al. 2008]{cha08} Chandra, P., Chevalier, R., \& Patat, F. 2008, ATel, 1391, 1
\bibitem[Chevalier 1982a]{che82a} Chevalier, R.~A. 1982a, ApJ, 258, 790
\bibitem[Chevalier 1982b]{che82b} Chevalier, R.~A. 1982b, ApJ, 259, 302
\bibitem[Chiotellis 2012]{chio12} Chiotellis, A., Schure, K.~M., \& Vink J. 2012, A\&A, 537, 139
\bibitem[Chomiuk et al.(2012)]{2012ApJ...750..164C} Chomiuk, L., Soderberg,
      A.~M., Moe, M., et al.\ 2012, ApJ, 750, 164 
\bibitem[Chornock et al.\ 2000]{cho00} Chornock, R. et al. 2000, IAU Circ., 7463
\bibitem[Chugai (2008)]{chu08} Chugai, N. 2008, Astronomy Letters, 34, 389
\bibitem[Chugai et al. 2004]{chuetal04} Chugai, N.~N., Chevalier, R.~A., \&
     Lundqvist, P. 2004, MNRAS, 355, 627
\bibitem[Chugai \& Yungelson (2004)]{chuyun04} Chugai, N.~N., \& Yungelson,
     L.~R. 2004, Astr. Letters, 30, 65
\bibitem[Crotts \& Yourdon(2008)]{2008ApJ...689.1186C} Crotts, A.~P.~S., \& Yourdon, D.\ 2008, ApJ, 689, 1186 
\bibitem[Cumming et al.\ 1996]{cum96} Cumming, R.~J., Lundqvist, P., 
     Smith, L.~J., Pettini, M., \& King, D.~L. 1996, MNRAS, 283, 
     1355 (C96)
\bibitem[Deng et al. (2004)]{Deng04} Deng, J., Kawabata, K.~S., Ohyama, Y., Nomoto, K., Mazzali, P.~A., Wang, L., Jeffery, D.~J., Iye, M., Tomita, H., \& Yoshii, Y. 2004, ApJ, 605, 37 
\bibitem[Dilday et al.(2012)]{2012Sci...337..942D} Dilday, B., Howell, D.~A., Cenko, S.~B., et al.\ 2012, 
Science, 337, 942
\bibitem[Di Stefano et al. (2011)]{Dist11} Di Stefano, R., Voss, R., \& Clayes, J.~S.~W. 2011, 
ApJ, 738, L1
\bibitem[D'Odorico \& Kaper 2000]{dod00} D'Odorico, S., \& Kaper, L. 2000, 
     UVES Users Manual
\bibitem[Dwarkadas \& Chevalier 1998]{dc98} Dwarkadas, V.~V., \& Chevalier, R.~A. 1998, ApJ, 497, 807
\bibitem[Eck et al.(2002)]{Eck02} Eck, C.~R., Cowan, J.~J., \& Branch, D. 2002, ApJ, 573, 306
\bibitem[Emsellem et al.\ 2004]{ems04} Emsellem, E., Cappellari, M., Peletier, R. F. et al. 2004, MNRAS, 352, 721
\bibitem[Fransson 1984]{f84} Fransson, C. 1984, A\&A, 133, 264
\bibitem[Fransson \& Bj\"ornsson 1998]{fb98} Fransson, C., \&
      Bj\"ornsson, C.-I. 1998, ApJ, 509, 861
\bibitem[Fransson et al.\ 1996]{flc96} Fransson, C., Lundqvist, P., \&
      Chevalier R.~A. 1996, ApJ, 461, 993
\bibitem[Gerardy et al.\ 2003]{ger03} Gerardy, C.~L. et al. 2003, ApJ, 607, 391
\bibitem[Hachisu et al.(1999a)]{Hach99a} Hachisu,  I., Kato, M., \& Nomoto, K. 1999a, ApJ, 522, 487
\bibitem[Hachisu et al.(1999b)]{Hach99b} Hachisu, I., Kato, M., Nomoto, K, \& Umeda, H. 1999b, 519, 314
\bibitem[Hachisu et al.(2008)]{Hach08} Hachisu, I., Kato, M., \& Nomoto, K. 2008, ApJ, 679, 1390
\bibitem[Hamuy et al.(2003)]{2003Natur.424..651H} Hamuy, M. et al., 2003, Nature, 424, 651
\bibitem[Han \& Podsiadlowski (2003)]{han06} Han, Z., \& Podsiadlowski, Ph. 2006, MNRAS, 368, 1095
\bibitem[Hillebrandt \& Niemeyer 2000]{HN00} Hillebrandt, W., \& Niemeyer, J.
     2000, ARA\&A, 38, 191
\bibitem[Hoyle \& Fowler 1960]{hf60}Hoyle, F., \& Fowler, W.~A. 1960,
     ApJ, 132, 565
\bibitem[Hughes et al.(2007)]{2007ApJ...670.1260H} Hughes, J.~P., Chugai, 
N., Chevalier, R., Lundqvist, P., \& Schlegel, E.\ 2007, ApJ, 670, 1260 
\bibitem[Iben \& Renzini 1983]{ir83}Iben, I. Jr., \& Renzini, A. 1983,
     ARA\&A, 21, 271
\bibitem[Iben \& Tutukov]{it84}Iben, I. Jr. \& Tutukov, A.~V. 1984,
     ApJS, 54, 335
\bibitem[Justham (2011)]{Just11} Justham, S. 2011, ApJ, 730, L34
\bibitem[Kasen(2010)]{2010ApJ...708.1025K} Kasen, D.\ 2010, ApJ, 708, 1025
\bibitem[Kasen et al.(2003)]{2003ApJ...593..788K} Kasen, D. et al. 2003, ApJ, 593, 788
\bibitem[Kotak et al. 2004]{ko04} Kotak, R., Meikle, W.~P.~S., Adamson, 
     A., \& Leggett, S. K. 2004, MNRAS, 354, L13
\bibitem[Kozma \& Fransson 1998]{kf98} Kozma, C., \& Fransson, C.. 1998, ApJ,
     496, 946
\bibitem[Kozma et al.\ 2005]{kzma05} Kozma, C., et. al. 2005, A\&A, 437, 983
\bibitem[Leonard et al.\ 2000]{leo00} Leonard, D.~C., Filippenko, A.~V., 
     Chornock, R., \& Li, W.~D. 2000, IAU Circ., 7471
\bibitem[Leonard(2007)]{2007ApJ...670.1275L} Leonard, D.~C.\ 2007, ApJ, 
670, 1275 
\bibitem[Li et al.\ 2001]{li01} Li, W.~D. et al., 2001, PASP, 113, 1178
\bibitem[Li et al.(2001)]{2001ApJ...546..734L} Li, W. et al. 2001, ApJ, 546, 734
\bibitem[Li et al.(2011)]{li11} Li, W., et al. 2011, Nature, 480, 348 
\bibitem[Liu et al.(2012)]{2012A&A...548A...2L} Liu, Z.~W., Pakmor, R., R{\"o}pke, F.~K., et al.\ 2012, A\&A, 548, A2 
\bibitem[Livio \& Riess (2003)]{liv03} Livio, M., \& Riess, A.~G. 2003, ApJ, 594, L93
\bibitem[Luna 2008]{luna08} Luna, R., Cox, N.~L.~J., Satorre, M.~A., Garcia Hernandez, D.~A., Suarez, O. \& Garcia Lario, P. 2008, A\&A, 480, 133
\bibitem[Lundqvist \& Cumming 1997]{lun97} Lundqvist, P., \& Cumming, R.~J. 
     1997, in Advances in Stellar Evolution, ed. R.~T. Rood \& A. Renzini 
     (CUP: Cambridge), 293 (LC97)
\bibitem[Lundqvist \& Fransson (1996)]{lf96} Lundqvist, P., \& Fransson, C.
     1996, ApJ, 464, 924
\bibitem[Lundqvist et al.(2005)]{lms05} Lundqvist, P. et al. 2004, in Proceedings ``Supernovae'', 
     IAU Coll. 192, eds. J.M. Marcaide \& K.W. Weiler (astro-ph/0309006), 
     CDROM p. 81
\bibitem[Lundqvist et al.(2003)]{2003fthp.conf..309L} Lundqvist, P. et al., 2003, in From Twilight to Highlight: 
     The Physics of Supernovae, eds. W. Hillebrandt \& B. Leibundgut, p. 309
\bibitem[Marietta et al.\ 2000]{mar00} Marietta, E., Burrows, A., \& Fryxell, B. 2000, ApJS, 128, 615
\bibitem[Mattila et al. (2005)]{Matt05b} Mattila, S. et al., 2005, A\&A, 443, 649 (M05)
\bibitem[Mattila et al. (2008)]{Matt08} Mattila S., Meikle W. P. S., Lundqvist P., et al. 2008, MNRAS, 389, 141
\bibitem[Mazzali et al.(2005)]{2005ApJ...623L..37M} Mazzali, P.~A., et al.\ 
2005, ApJl, 623, L37 
\bibitem[Mihalas (1978)]{Mih78} Mihalas, D. 1978, Stellar Atmospheres (2nd edition), San Francisco, W. H. Freeman and Co.
\bibitem[Munari et al.\ 1998]{mun98} Munari, U., Barbon, R., Tomasella, L.,
     \& Rejkuba, M. 1998, IAU Circ., 6902
\bibitem[Munari \& Zwitter 1997]{mz97} Munari, U., \& Zwitter, T. 1997, 
     A\&A, 318, 269
\bibitem[Nomoto et al.\ 1999]{nom99} Nomoto, K., Nakamura, T., \& Kobayashi, 
     C. 1999, AP\&SS, 265, 37
\bibitem[Nomoto et al.\ 1984]{nom84} Nomoto, K., Thielemann, F.-K., \& 
     Yokoi, K. 1984, ApJ, 286, 644
\bibitem[Paczy\'nski 1985]{p85} Paczy\'nski, B. 1985, in Cataclysmic Variables
     and Low-Mass X-Ray Binaries, ed. D.~Q. Lamb \& J. Patterson (Dordrecht:
     Reidel), p. 1
\bibitem[Pakmor et al.(2008)]{2008A&A...489..943P} Pakmor, R., R{\"o}pke, F.~K., Weiss, A., \& Hillebrandt, W.\ 
     2008, A\&A, 489, 943 
\bibitem[Pan et al.(2012)]{2012ApJ...750..151P} Pan, K.-C., Ricker, P.~M.,
     \& Taam, R.~E.\ 2012, ApJ, 750, 151 
\bibitem[Patat et al.(2006)]{2006MNRAS.369.1949P} Patat, F., Benetti, S., Cappellaro, E., \& Turatto, M.\ 2006,
     MNRAS, 369, 1949 
\bibitem[Patat et al.(2007a)]{2007A&A...474..931P} Patat, F. et al.\ 2007, A\&A, 474, 931
\bibitem[Patat et al.(2007b)]{2007Sci...317..924P} Patat, F. et al.\ 2007a, Science, 317, 924
\bibitem[Patat etal.(2011)]{2011A&A...530A..63P} Patat, F., Chugai, N.~N., Podsiadlowski, P., 
     et al.\ 2011, A\&A, 530, A63  
\bibitem[Perlmutter et al.\ 1999]{perl99} Perlmutter, S. et al. 1999, ApJ, 517, 565
\bibitem[Prieto et al. (2005)]{Pri05} Prieto, J., et al. 2005, IAU Circ. 8633
\bibitem[Riess et al.(1999)]{1999AJ....118.2675R} Riess, A.~G. et al. 1999, AJ, 118, 2675
\bibitem[Ruiz-Lapuente et al.\ 2004]{rl04} Ruiz-Lapuente, P. et al. 2004, Nature, 431, 1069
\bibitem[Schaefer \& Pagnotta 2012]{scha12} Schaefer, B.~E., \& Pagnotta, A. 2012, Nature 481, 164
\bibitem[Schlegel et al.\ 1998]{sch98} Schlegel, D.~J., Finkbeiner, D.~P.,
     Davis, M. 1998, ApJ, 500, 525
\bibitem[Schlegel \& Petre 1993]{schp93} Schlegel, E.~M., \& Petre, R. 
     1993, ApJ, 418, L53
\bibitem[Schmidt et al.\ 1998]{schm98} Schmidt, B.~P. et al. 1998 ApJ, 507, 46
\bibitem[Shappee et al. (2013a)]{Sha13} Shappee, B.~J., Kochanek, C.~S., \& Stanek, K.~Z. 2013a, ApJ, 765, 150
\bibitem[Shappee et al.(2013b)]{2013ApJ...762L...5S} Shappee, B.~J., Stanek,
     K.~Z., Pogge, R.~W., \& Garnavich, P.~M.\ 2013b, ApJ, 762, L5 
\bibitem[Silverman et al.(2013)]{2013arXiv1304.0763S} Silverman, J.~M.,
     Nugent, P.~E., Gal-Yam, A., et al.\ 2013, arXiv:1304.0763 
\bibitem[Simien \& Prugniel 2000]{sp00} Simien \& Prugniel, 2000, A\&AS, 145, 263
\bibitem[Soker et al. (2013)]{Sok13} Soker, N., Kashi, A., Garcia-Berro, E., Torres, S., \& Camacho,
J., MNRAS, 431, 1541
\bibitem[Sollerman et al. (2004)]{Sol04} Sollerman, J. et al. 2004, A\&A, 428, 555
\bibitem[Sollerman et al. (2005)]{Sol05} Sollerman, J. et al. 2005, A\&A, 429, 559
\bibitem[Sorokina et al. (2004)]{Sor04} Sorokina E. I., Blinnikov S. I., Kosenko D. I., Lundqvist P. 2004, 
      Astron. Lett., 30, 737
\bibitem[Spyromilio et al. (2004)]{Spy04} Spyromilio, J. 2004 A\&A, 426, 547
\bibitem[Stanishev et al.(2007)]{2007A&A...469..645S} Stanishev, V., et al.\ 2007, A\&A, 469, 645 
\bibitem[Sternberg et al. (2011)]{Ste11} Sternberg, A., et al. 2011, Science, 333, 856
\bibitem[Taam 1980]{T80} Taam, R. 1980 ApJ, 237, 142
\bibitem[Thielemann et al.\ 1986]{thiel86} Thielemann, F.-K., Nomoto, K., \& 
     Yokoi, K. 1986, A\&A, 158, 17
\bibitem[Tanaka et al.(2008)]{2008ApJ...677..448T} Tanaka, M., et al.\ 2008, ApJ, 677, 448 
\bibitem[Thomas et al.(2004)]{2004ApJ...601.1019T} Thomas, R.~C. et al. 2004, ApJ, 601, 1019
\bibitem[Trundle et al. (2008)]{Tru08} Trundle, C., Kotak, R., Vink, J.~S, \& Meikle, W.~P.~S. 2008, A\&A, 483, L47
\bibitem[Wang et al.(2008)]{2008ApJ...677.1060W} Wang, X., Li, W., Filippenko, A.~V., Foley, R.~J., 
      Smith, N., \& Wang, L.\ 2008, ApJ, 677, 1060
      395, 847  
\bibitem[Webbink 1984]{W84} Webbink, R.~F. 1984, ApJ, 277, 355
\bibitem[Whelan \& Iben(1973)]{WI73} Whelan, J. \& Iben, I., Jr. 1973, ApJ,
      186, 1007
\end{thebibliography}
\end{document}